\documentclass[journal]{IEEEtran}
\usepackage{mathrsfs}
\usepackage{amsmath}
\usepackage{amssymb}
\usepackage{mathtools} 
\usepackage{amsthm}
\usepackage{mathptmx}
\DeclareMathAlphabet{\mathcal}{OMS}{cmsy}{m}{n}
\DeclareSymbolFont{largesymbols}{OMX}{cmex}{m}{n}
\usepackage{verbatim}
\usepackage{amsfonts}
\usepackage{cite}
\usepackage{algorithm}
\usepackage{cuted}
\usepackage{lipsum}
\usepackage{algpseudocode}
\usepackage{booktabs}
\usepackage{amstext}
\usepackage{bm}
\usepackage[pdftex,linkcolor=blue,citecolor=blue]{hyperref}
\hypersetup{hidelinks}

\usepackage[hyphenbreaks]{breakurl}
\usepackage{threeparttable}
\usepackage{nomencl}
\usepackage{float}
\makenomenclature

\usepackage{url}

\allowdisplaybreaks[4]

\usepackage{float}

\usepackage[pdftex]{graphicx}
\graphicspath{{../pdf/}{../jpeg/}}
\DeclareGraphicsExtensions{.pdf,.jpeg,.png}

\newcommand{\RNum}[1]{\uppercase\expandafter{\romannumeral #1\relax}}

\def\baa{\begin{align}}
\def\eaa{\end{align}}

\newcommand{\bsq}{\begin{subequations}}
	\newcommand{\esq}{\end{subequations}}

\newcommand{\beq}{\begin{equation}}
\newcommand{\eeq}{\end{equation}}
\newcommand{\bq}{\begin{eqnarray}}
\newcommand{\eq}{\end{eqnarray}}
\newcommand{\bqn}{\begin{eqnarray*}}
	\newcommand{\eqn}{\end{eqnarray*}}
\newcommand{\bee}{\begin{enumerate}}
	\newcommand{\eee}{\end{enumerate}}
\newcommand{\bi}{\begin{itemize}}
	\newcommand{\ei}{\end{itemize}}

\usepackage{comment}
\usepackage{color}
\newboolean{showcomments}
\setboolean{showcomments}{true}
\newcommand{\wang}[1]{\ifthenelse{\boolean{showcomments}}
	{ \textcolor[rgb]{1,0,1}{(ZW:  #1)}}{}}
\newcommand{\fliu}[1]{\ifthenelse{\boolean{showcomments}}
	{ \textcolor{red}{(FL:  #1)}}{}}
\newcommand{\zhang}[1]{\ifthenelse{\boolean{showcomments}}
	{ \textcolor{blue}{(YFZ:  #1)}}{}}

\theoremstyle{definition}
\newtheorem{theorem}{Theorem}
\newtheorem{lemma}[theorem]{Lemma}
\newtheorem{corollary}[theorem]{Corollary}

\theoremstyle{definition}
\newtheorem{definition}{Definition}
\newtheorem{remark}{Remark}

\newtheorem{example}{Example}
\newtheorem{assumption}{\textit{Assumption}}

\ifCLASSOPTIONcompsoc
\usepackage[caption=false, font=normalsize, labelfont=sf, textfont=sf]{subfig}
\else
\usepackage[caption=false, font=footnotesize]{subfig}
\begin{document}

\title{On Decision-Dependent Uncertainties in Power Systems with High-Share Renewables}

\author{Yunfan Zhang,
		Yifan Su,
        Feng Liu

	\thanks{This work was supported by the Joint Research Fund in Smart Grid (No.U1966601) under cooperative agreement between the National Natural Science Foundation of China (NSFC) and State Grid Corporation of China. \textit{(Corresponding author: Feng Liu)}}
}

\maketitle

\begin{abstract}
The continuously increasing renewable energy sources (RES) and demand response (DR) are becoming important sources of system flexibility. As a consequence, decision-dependent uncertainties (DDUs), interchangeably referred to as endogenous uncertainties, impose new characteristics to power system dispatch. The DDUs faced by system operators originate from uncertain dispatchable resources such as RES units or DR, while reserve providers encounter DDUs arising from the uncertain reserve deployment. This paper presents a systematic framework for addressing robust dispatch problems with DDUs. The main contributions include i) the robust characterization of DDUs with a dependency decomposition structure; ii) a generic DDU coping mechanism, manifested as the bilateral matching between uncertainty and flexibility; iii) analyses of the influence of DDU incorporation on the convexity/non-convexity of robust dispatch problems; and iv) generic solution algorithms adaptive for DDUs. Under this framework, the inherent distinctions and correlations between DDUs and DIUs are revealed, providing a fundamental theoretical basis for the economic and reliable operation of RES-dominated power systems. Applications in the source and demand sides illustrate the importance of considering DDUs and verify the effectiveness of proposed algorithms for robust dispatch with DDUs.
\end{abstract}


\IEEEpeerreviewmaketitle

\section{Introduction}
\subsection{Background}
Driven by the imperative to reduce greenhouse gas emissions and mitigate the impacts of climate change, recent years have witnessed a worldwide unprecedented proliferation of renewable energy sources (RES) in power systems. 

Alongside the remarkable benefits of the transition towards a RES-dominated power system, the inherent variability and intermittency of RES, notably the photovoltaic (PV) and wind generations, exacerbates the variable characteristic of the power system and arises significant challenges to ensure the reliability and economic efficiency of system operation. Various uncertainty handling methods, including stochastic optimization\cite{chen2012stochastic}, robust optimization (RO)\cite{jiang2012robust}, distributionally robust optimization (DRO)\cite{xiong2017adistributionally}, the information gap decision theory (IGDT)\cite{majidi2019application}, etc., have been studied to hedge against the uncertainties in power systems. Due to the high dispatch decision robustness and low requirement in uncertainty modeling, RO has been widely adopted in real power systems. PJM and Alstom Grid have developed the two-stage robust optimization since 2012 \cite{wang2013two}. CAISO has also factored in heterogeneous uncertainties into its optimization framework through RO \cite{CAISO}. There are still some challenges on heavy computational burden and trade-off between robustness and performance, which, however, out of scope for this paper.


These optimization methods distinguish themselves in uncertainty characterization and risk management techniques. Moreover, the considered uncertain factors stemming from the inaccurate prediction of RES generations or load demand, the disruptions of network or generation units, and the variable market clearing prices, are typically treated as exogenous variates independent of the decision-making outcome.

Decision-dependent uncertainties (DDUs), also known as endogenous uncertainties, have received increasing attention recently due to the escalating RES penetration and the emerging novel features of power systems. In real-world decision problems, uncertainties could be affected by decisions in three possible ways\cite{zhang2020aunified}: 
\begin{itemize}
	\item Decisions alter the underlying probability distribution of the random variable. For example, in a network subject to random failures, the survival probability of a link is increased if it is strengthened by investment \cite{Kannan2004investing}.
	\item Decisions affect whether or when the random variable materializes, i.e., retains physical meaning. For example, the uncertain output of a wind farm will materialize only if the decision is made to construct it.
	\item Decisions affect the resolution of the uncertainty, i.e., whether or when the uncertain variable can be observed by the decision maker. For example, the product demand curve is fixed but unknown to the retailer but he can exploit and explore the characteristics of customer behavior by different pricing decisions\cite{bertsimas2014datadriven}.
\end{itemize}
Contrary to DDUs, uncertainties unaffected by decisions are termed decision-independent uncertainties (DIUs) or exogenous uncertainties interchangeably.

\subsection{DDU in Power System Dispatch}
DDUs in the dispatch problems of power systems typically arise from three aspects: i) the uncertain dispatchable resources on the source side; ii) the uncertain dispatchable resources on the demand side; and iii) the uncertain reserve deployment from the operators.

{\textbf{i) DDUs Originating from Uncertain Dispatchable Resources: the Source Side.}} In high-RES-penetration power systems, besides conventional synchronous generators, the RES units can serve as dispatchable resources.
By incorporating frequency-related signals within the wind turbine's active power control loop, the wind turbine can be responsive to frequency variations, which helps to stabilize the system deviations. However, the kinetic energy stored in the rotating mass is limited and the discharge of energy to the grid is only available for a short period. 
To participate in primary, secondary, and tertiary frequency control, the wind turbine needs to retain a certain amount of spinning power reserve by deviating from the maximum power point (MPP) through rotor speed or pitch angle control, to support system balance in case of frequency contingency.
Similarly, PV generations can be curtailed below MPP through voltage control to provide reserve power.
Alternatively, the output of a wind farm or a PV plant can be fast regulated through controlled shut-down or start-up of the units.

\begin{figure*}[!htb]
	\centering
	\subfloat[The available power in \textit{De-rating} scheme.]{\label{fig:chp2:deload:b}
		\includegraphics[width=0.3\linewidth]{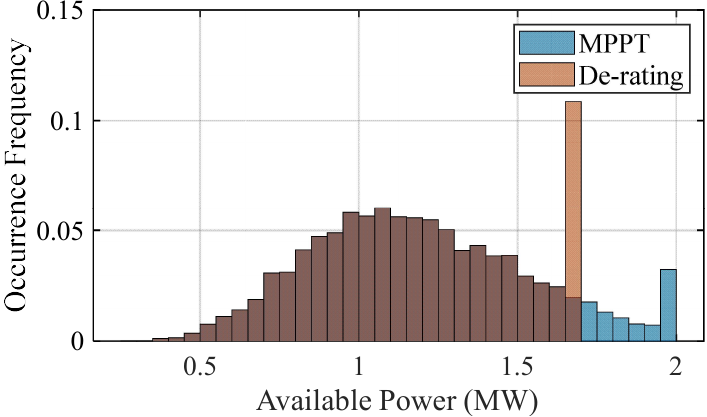}}
	\subfloat[The available power in \textit{Delta} scheme.]{\label{fig:chp2:deload:c}
		\includegraphics[width=0.3\linewidth]{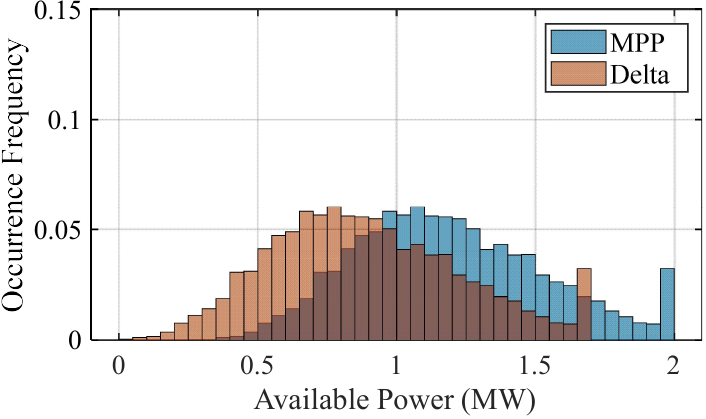}} 
	\subfloat[The available power in \textit{Percentage} scheme]{\label{fig:chp2:deload:d}
		\includegraphics[width=0.3\linewidth]{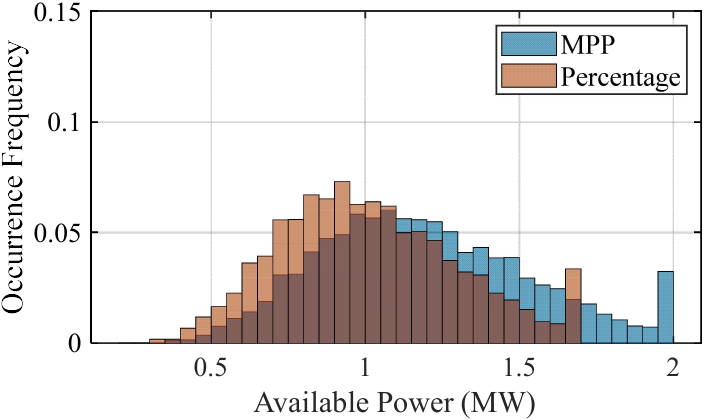}}\\
	\subfloat[The preserved power in \textit{De-rating} scheme.]{\label{fig:chp2:deload:e}
		\includegraphics[width=0.3\linewidth]{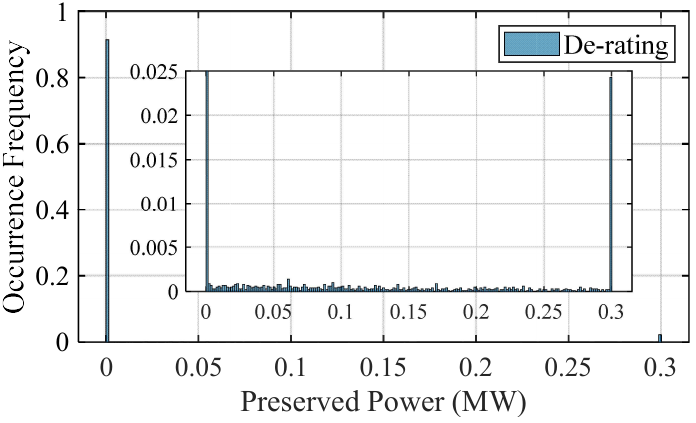}}
	\subfloat[The preserved power in \textit{Delta} scheme.]{\label{fig:chp2:deload:f}
		\includegraphics[width=0.3\linewidth]{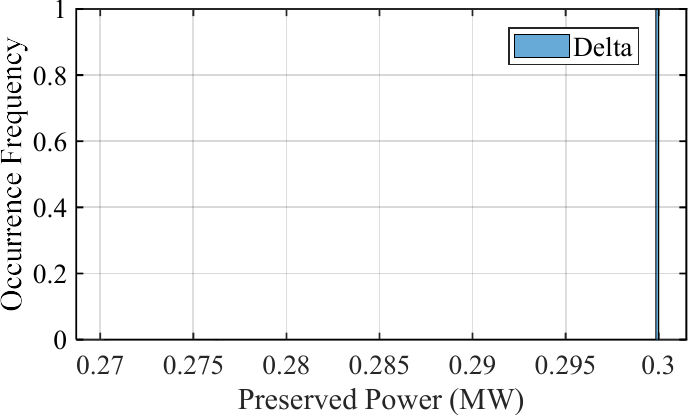}} 
	\subfloat[The preserved power in \textit{Percentage} scheme]{\label{fig:chp2:deload:g}
		\includegraphics[width=0.3\linewidth]{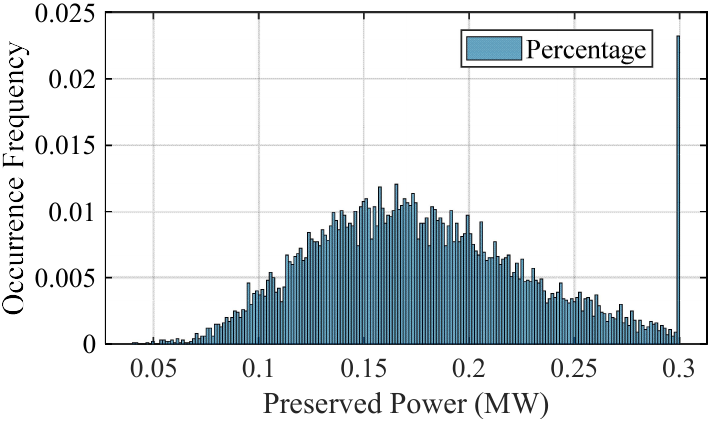}}
	\caption{Histogram of the sampled available power and preserved power of a wind turbine with the rated power of 2MW. The cut-in, rated, and cut-off wind speed is set as 4 m/s, 10 m/s, and 22 m/s, respectively. The wind speed data are sampled through a normal distribution with a mean of 9 m/s. The reserve ratio is set as $\lambda=0.15$.}
	\label{fig:chp2:deload} 
\end{figure*}

It is worth noting that, the regulation effects of RES are jointly determined by the variable weather conditions and the proactive power control strategies or actions of the operators. Therefore, the available power and the reserved power of RES units manifest as DDUs. To exemplify the RES-induced DDUs, an illustration example is provided: The RES units typically operate in three de-loading schemes\cite{li2021frequency,fleming2016effects,mypaperfreq} to provide sustainable up-reserve for frequency regulations where the decision variable $0\le\lambda\le 1$ denotes the de-loading ratio: 
\begin{itemize}
	\item The \textit{De-rating} scheme: The maximal output of the RES unit is capped at a de-rated point, which equals $1-\lambda$ of the rated power. \item The \textit{Delta} scheme: Setting the reserve into a fixed amount that equals $\lambda$ of the rated power.
	\item The \textit{Percentage} scheme: The RES unit reserves a percentage ($\lambda$) of the maximum available power.
\end{itemize}
It is observed in Fig.\ref{fig:chp2:deload} that, the available and preserved power of a wind turbine is uncertain due to the volatile wind speed; and decisions of the system operator, including the de-loading scheme and the reserve ratio $\lambda$, do affect the probability distribution characteristics of these two uncertain parameters.

The DDUs of RES units have attracted increasing concerns in the literature. 
In \cite{wang2016robust,chen2022robust}, a day-ahead strategic wind curtailment schedule is made to reduce wind variability and the two-stage robust dispatch model is built to address the corresponding DDU of 
the real-time wind generation output. 
In \cite{aigner2021robust}, DDUs arising from the solar feed-in curtailment are dealt with in a jointly chance-constrained stochastic DC optimal power flow model, where affine decision rules are applied to adjust generator output once the uncertainties are revealed.
Reference \cite{li2021frequency} presents a  frequency-constrained stochastic planning model with frequency response from de-loaded wind farms, where the support capability uncertainty and its dependency upon the combination of different support schemes are considered. In \cite{mypaperfreq}, the DDUs arising from frequency reserve allocation among RES units are addressed without assuming a fixed portion of power reserve used for the synthetic inertial response and the droop response when modeling the post-contingency system frequency dynamics.
 
{\textbf{ii) DDUs Originating from Uncertain Dispatchable Resources: the Demand Side.}} In the context of demand response (DR), customers respond to price signals or incentives, adjusting their electricity consumption patterns to contribute flexibility to system balance. The actual DR behaviors of voluntary customers are inevitably uncertain while closely related to price signals or other DR policies, therefore manifesting as DDUs. The exploration of DDUs in DR would be reviewed below, encompassing both price-elastic models \cite{zhao2013multistage,wang2013stochastic,wang2021optimal,liu2015robust} and non-price models\cite{su2020robust,su2022multistage,li2022restoration,giannelos2018option}.

\begin{figure}[!htb]
	\centering
	\includegraphics[width=0.6\linewidth]{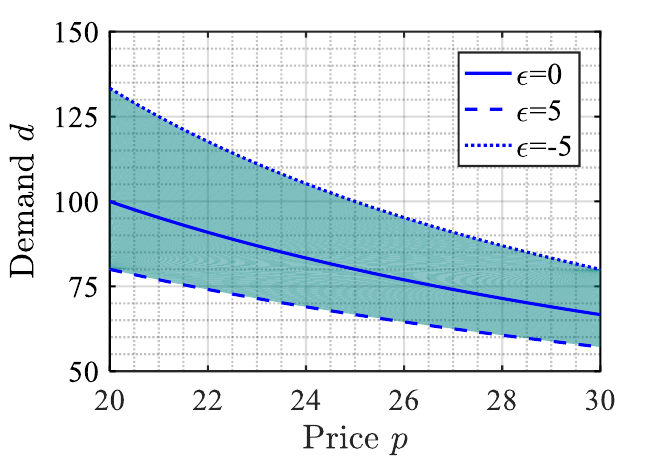}
	\caption{The uncertain price-elastic demand curve \cite{zhao2013multistage}. The formulation is $d=A(p+\epsilon)^{\alpha}$ where $d$ is the electricity demand, $A$ and $\alpha$ are known coefficients, $p$ denotes the price, and $\epsilon$ represents the random deviation that varies within $-\hat{\epsilon}\le \epsilon \le\hat{\epsilon}$. Set $A=2000$, $\alpha=-1$, and $\hat{\epsilon}=5$ \cite{zhao2013multistage}, then characteristics of the DDU variable $d$ are plotted.}
	\label{fig:dr:b}
\end{figure}

\begin{figure}[!htb]
	\centering
	\includegraphics[width=0.6\linewidth]{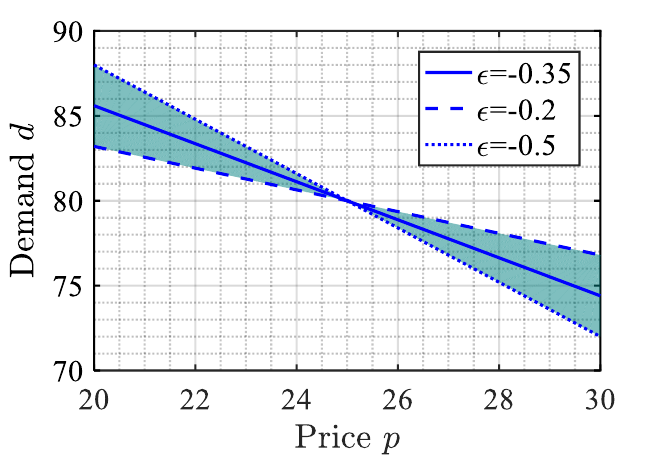}
	\caption{The uncertain price-elastic demand curve \cite{liu2015robust}. Denote by $d$ the uncertain demand and $d_0$ the base load without any DR. Denote by $p$ the price corresponding to $d$ and $p_0$ the reference price when demand $d_0$ is not reduced or increased. The price-elastic demand curve is formulated as  
	$\frac{d-d_0}{d_0}=\epsilon\times \frac{p-p_0}{p_0}$
	where the price-elasticity coefficient $\epsilon$ is a random variable that falls within $[\epsilon^{\rm min},\epsilon^{\rm max}]$. 
	Set $p_0=25$ \$/MWh, $d_0=80$ MWh, $\epsilon^{\rm min}=-0.5$, and $\epsilon^{\rm max}=-0.2$, then characteristics of the DDU variable $d$ are plotted.}
	\label{fig:dr:c}
\end{figure}

\begin{figure}[!htb]
	\centering
	\includegraphics[width=0.9\linewidth]{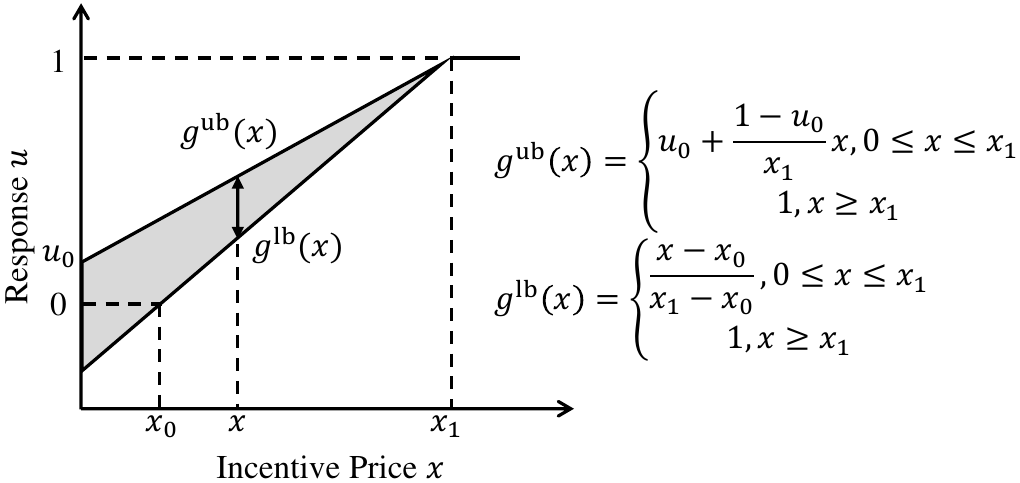}
	\caption{The uncertain user participation in DR\cite{wang2021optimal}.
	The uncertain user DR response is denoted by $u$ and the price incentive is $x$. Characteristics of the DDU variable $u$ are plotted.}
	\label{fig:chp2:dr:data:two}
\end{figure}

The price-elastic models: In \cite{zhao2013multistage}, the uncertain price-elastic demand curve is integrated into a robust unit commitment problem. A deviation term is introduced to the demand-price function, see Fig. \ref{fig:dr:b} for illustration. As indicated in the figure, for a given price, the consumer response varies within a certain range, with this range dependent on the price decision. 
Reference \cite{wang2013stochastic} represents the uncertainty of the price-elastic demand curve through scenarios in the stochastic unit commitment problem, with each scenario corresponding to a curve.
Reference \cite{liu2015robust} assumes that the uncertain price elasticity coefficients vary within a specified range, leading to a price-sensitive range within which the actual load fluctuates, as depicted in Fig. \ref{fig:dr:c}. In \cite{wang2021optimal}, the DR uncertainty model is established considering consumer psychology: as the incentive price rises, the willingness of users to participate in DR would increase, while the associated randomness of the participation would decrease accordingly. An illustration example is provided in Fig. \ref{fig:chp2:dr:data:two}.


The non-price models: Reference \cite{su2020robust} proposes a two-stage robust economic dispatch model wherein the DR setpoint determined by the system operator influences the high-dimensional box-type decision-dependent uncertainty set of the load demand. In the subsequent study \cite{su2022multistage}, DDUs of deferrable loads are modeled explicitly, as the cross-time load shifting decisions would reshape the temporal distribution of load and the corresponding uncertainty set. Consequently, a multi-stage robust dispatch model is formulated, accompanied by the design of an enhanced constraint generation algorithm. Reference \cite{li2022restoration} focuses on DDUs stemming from the cold load pickup (CLPU) phenomenon after an outage and develops a two-stage stochastic restoration model. Different load pickup time results in distinguished probability distributions of the CLPU peak amplitude. In \cite{giannelos2018option}, DDUs emerge as the uncertain consumer participation in a DR program can only be resolved upon actual deployment of the DR scheme. To address this, a scenario-wise multi-stage stochastic planning model is proposed, incorporating DDUs through the non-anticipativity constraints in the scenario tree.

{\textbf{iii) DDUs Originating from Uncertain Reserve Deployment.}} In the reserve market or energy-reserve market, participants autonomously decide on their reserve capacity provision through strategic bidding \cite{mashhour2011bidding1,mashhour2011bidding2} or price-based self-scheduling \cite{Dabbagh2015risk,baringo2019dayahead}. The actual deployment of reserve power during real-time operation is uncertain, contingent upon the occurrence of disturbances and regulation signals from the operator. However, the uncertain deployed reserve power must fall within the capacity available for dispatch, thus manifesting as DDUs for reserve providers. Reference \cite{zhang2017robust} presents an example of commercial buildings providing frequency reserves to the power grid. To determine the optimal offered reserve capacity, a robust optimal control problem with an adjustable uncertainty set is formulated. Reference \cite{my2022robust} addresses the robust self-scheduling of a virtual power plant participating in the day-ahead energy-reserve market, wherein DDUs pertaining to real-time reserve deployment requests are captured through a polyhedral uncertainty set with decision-dependent right-hand-side vector.

\begin{remark}[Time Scale of DDU]
DDU exists in optimization problems of power systems across different time scales. References \cite{wang2016robust,chen2022robust} make \textit{day-ahead} decisions of wind curtailment, which leads to decision-dependent wind variability. In reference \cite{su2022multistage}, deferrable loads are shifted across hours, causing \textit{intra-day} DDU of power demand. Besides, in the field of power system planning, \textit{monthly and yearly} time-scale DDUs of wind generation are commonly considered \cite{yin2022coordinated, yin2022stochastic}. In this paper, the proposed theorems and algorithms can be applied to problems of any time scale. Therefore, the following sections of this paper will not emphasize specific time scales.
\end{remark}

\subsection{Contributions}
While much has been written about the DDUs in power systems, more systematic theoretical researches on this topic are needed. In our prior work \cite{mypaper2022two}, the solution techniques for DDU-integrated RO have been reviewed, encompassing both the static RO models \cite{poss2013robust,poss2013robust2,nohadani2018optimization,lappas2018robust,lappas2017theuse} and the two-/multi-stage RO (TSRO) models \cite{lappas2016multistage,feng2020multistage,zhang2020aunified,avraamidou2020adjustable,mypaper2022two}. The previous studies collectively indicate that, DDUs would lead to more sophisticated optimization models, posing challenges in developing exact and efficient solution algorithms. However, the underlying reasons for these challenges and common solutions have not been thoroughly explored, to the best of our knowledge.

In this work, we develop a systematic framework for two-stage robust dispatch with DDUs. The fundamental differences and relationships between DDUs and DIUs, from the aspects of uncertainty characterization, coping mechanisms, problem solvability, and solution paradigms, are explicitly investigated. 

The main contributions of this paper are as follows:


\textbf{i) Robust characterization of DDUs:} The distinctions and correlations between DDUs and DIUs are elucidated within the framework of set-based uncertainty characterization and TSRO dispatch. A novel concept, the separability of DDUS, is introduced to articulate the dependency structure between decision variables and random variables. The existence of the separability is further investigated, showcasing that any generic DDUS can be decomposed into a decision-independent support, pertaining to the robust formulation of uncertainty, and a coupling function that describes the dependency on decisions.

\textbf{ii) Mechanism of TSRO dispatch with DDUs:} The mechanism of TSRO dispatch under DDUs is revealed within the framework of set-based uncertainty characterization and region-based flexibility characterization \cite{wu1982steady,wei2015dispatchable}. It is discovered that a TSRO dispatch problem essentially aims at obtaining the optimal allocation of system flexibility to hedge against uncertainties and DDUs enable the bi-directional matching between them. 

\textbf{iii) Convexity of TSRO dispatch with DDUs:} The effects of DDUs on the characteristics and solvability of TSRO dispatch problems are investigated. Reference \cite{wei2013phd} proves that for a TSRO problem that contains only linear constraints and DIUs, its robust feasibility region (RFR) is convex. This paper reveals that, in contrast to DIUs, the RFRs of TSRO dispatch problems with DDUs may be non-convex. The non-convexity, if any, is attributed to the dependency of the uncertainty set on decisions and is irrelative to the convexity of the uncertainty set itself. With the concept of separability, the convexity of the coupling function is proved to be a  sufficient condition for ensuring the convexity of the TSRO problem with DDUs.

\textbf{iv) Generic solution algorithm for TSRO dispatch with DDUs:} 
Traditional cutting-plane algorithms widely used to solve robust dispatch problems, such as Benders decomposition and the C\&CG algorithm, are not adapted to the non-convexity arising from DDUs. Their limitations in addressing TSRO with DDUs are evidenced by the ruined robust feasibility and optimality of the solution. To fix this issue, an improved solution algorithm is proposed based on the separability of DDUS. The convexity of the enhanced cuts therein automatically adapts to that of the TSRO problem.

\subsection{Organization}
The rest of this paper is organized as follows. In Section II, the dependency decomposition structure of DDUs in robust frameworks is revealed. Section III derives the mechanism of robust dispatch with DDUs, while Section IV discusses the convexity of robust dispatch problems with DDUs. In Section V, a general improved solution strategy is designed for DDU-integrated robust optimization problems. Section VI presents applications of robust dispatch with DDUs, while Section VII concludes this paper.

\section{The Robust Characterization of Decision-Dependent Uncertainties} \label{sec2}


Consider the TSRO model in \eqref{tsro}, which has been widely used as the dispatch model in many real systems. 
\begin{subequations}
	\label{tsro}
	\begin{gather}
		\label{tsro:1}
		\min\nolimits_{x}\quad f(x)+S(x)\\
		\label{tsro:2}
		\text{s.t.}\quad x\in X\cap X_R\\
		\label{tsro:3}
		S(x):=\max_{u\in \mathcal{U}(x)}\min_{y\in Y(x,u)}\ c^{\mathsf{T}}y\\
		\label{tsro:5}
		Y(x,u):= \left\{
		y\in\mathbb{R}^{n_y}\ |\ Ax+By+Cu\le b,y\ge \mathbf{0}
		\right\}\\
		\label{tsro:4}
		X_R:= \left\{x\ |\ 
		Y(x,u)\neq \emptyset,\forall u\in \mathcal{U}(x)\right\}
	\end{gather}
\end{subequations}
In \eqref{tsro}, $x$ denotes the here-and-now decision with constraint set $X$. $u$ is the DDU variable affected by $x$. The uncertainty set of $u$, denoted by $\mathcal{U}(x):X\rightrightarrows U$ is a set-valued map parameterized in $x$ and $U$ is a set that contains all possible realizations of $u$. According to the set-based uncertainty representing method in RO, for any given $x$, $\mathcal{U}(x)$ characterizes the possible realizations of $u$ under a certain level of confidence. $f(x)$ is the first-stage cost function and $S(x)$ which is the optimal value of a max-min optimization problem indicates the worst-case second-stage cost. The second-stage dispatch problem upon the resolution of the uncertainty is assumed to be a linear programming parameterized in $x$ and $u$ and with wait-and-see decision variable $y$, as shown in \eqref{tsro:3}-\eqref{tsro:5}. $Y(x,u)$ is the feasible region of $y$. $X_R$ is the RFR.

The following mild assumptions are made to problem \eqref{tsro}.
\begin{assumption}
\label{assp-tsro}
For problem \eqref{tsro},
\begin{itemize}
	\item[(a)] Set $X$ is a convex set.
	\item[(b)] Function $f(x):X\rightarrow \mathbb{R}^1$ is convex on $X$.
	\item[(c)] Set $U$ is a bounded polytope.
	\item[(d)] For any $x\in X$ and $u\in U$, $Y(x,u)$ is a bounded polytope.
\end{itemize}
\end{assumption}

\begin{remark}[Assumption on Polyhedral Model]
\textbf{\textit{Assumption}} \ref{assp-tsro} restricts that sets $U$ and $Y(x,u)$ are bounded polytopes, which is common in power system robust dispatch problems. Operational constraints of fuel or wind generators, energy storage, and other devices are mainly linear. Though power flow model is nonlinear and nonconvex, the linearized models, e.g., DC flow model for transmission networks \cite{stott2009dc} and LinDistFlow model for distribution networks \cite{zhu2015fast} have been widely applied. Therefore, \textbf{\textit{Assumption}} \ref{assp-tsro} is practical for TSRO dispatch in power systems.
\end{remark}

\begin{remark}[Tri- and Multi-Stage Robust Optimization with DDUs]
Though this paper focuses on two-stage robust optimization problems like \eqref{tsro}, the proposed theorems and algorithms can be adopted to tri- or multi-stage problems. C\&CG and Benders' decomposition algorithms for TSRO can be nested to solve tri-stage problems. For multi-stage one, affine policies are commonly utilized to simplify the models so that it can be solved by cutting-plane algorithms. Therefore, it is potential to extend the results in TSRO to tri- and multi-stage robust optimization with DDUs.
\end{remark}

Next, the dependency structure of DDUS $\mathcal{U}(x)$ is investigated by introducing a novel concept of separability.
\subsection{The Complete Separability of DDUS}
\label{subsec:complete_separa}

\subsubsection{Definition}
\begin{definition}[Complete Separability of DDUS]
\label{def:sep:comp}
If a DDUS $\mathcal{U}(x)$ can be equivalently rewritten into \eqref{chp2:def:seperate},
\begin{align}
\label{chp2:def:seperate}
\mathcal{U}(x)=\left\{
u=\mathcal{C}(\xi,x)\ |\ \xi\in \Xi
\right\},\forall x\in X
\end{align}
then $\mathcal{U}(x)$ is called (completely) \textbf{separable} where $\xi$ is the auxiliary random variable with decision-independent support $\Xi$ and $\mathcal{C}(\cdot):\Xi\times X \rightarrow U$ is the coupling function parameterized in $\xi$ and $x$. A DDUS $\mathcal{U}(x)$ with separable formulation like \eqref{chp2:def:seperate} is decomposed into two parts: the dependency on decision variable $x$ denoted by the coupling function $\mathcal{C}(\cdot)$ and the decision-irrelevant stochasticity implied by the random variable $\xi$ and its support $\Xi$. 
\end{definition}

\subsubsection{Ubiquity}


Next, we would like to show the ubiquity of the separability of DDUS by taking space $U$ as the decision-irrelevant support set. 

Without loss of generality, a DDUS $\mathcal{U}(x):X\rightrightarrows U$ can be denoted as the intersection of two sets:
\begin{gather}
	\label{chp2:ddus:1}
	\mathcal{U}(x)=U^{\rm sub}\cap \mathcal{U}^{\rm sup}(x),\forall x\in X
\end{gather}
where ${U}^{\rm sub}\subseteq U$ is a decision-independent set in space $U$ and $\mathcal{U}^{\rm sup}(x):X\rightrightarrows U$ is a set-valued map parameterized in decision $x$. Note that the intersecting form in \eqref{chp2:ddus:1} does not impose additional requirements onto the specific formulation of DDUS $\mathcal{U}(x)$, as one can always take ${U}^{\rm sub}$ as $U$ and $\mathcal{U}^{\rm sup}(x)$ as $\mathcal{U}(x)$. The following lemma holds for any DDUS with intersecting form as in \eqref{chp2:ddus:1}.

\begin{lemma}
\label{lemma:1}
Suppose for DDUS $\mathcal{U}(x):X\rightrightarrows U$ that can be denoted by the intersection of two sets as in \eqref{chp2:ddus:1}, there exists a coupling function $\mathcal{C}(u,x):{U}^{\rm sub}\times X\rightarrow U$ such that: 
\begin{itemize}
\item [(a)] for any $x\in X$ and $u\in\mathcal{U}(x)$, $\mathcal{C}(x,u)=u$;
\item [(b)] for any $x\in X$ and $u\in {U}^{\rm sub}$, $\mathcal{C}(u,x)\in \mathcal{U}(x)$. 
\end{itemize}
Then, $\mathcal{U}(x)$ is with complete separability as follows:
\begin{gather}
	\label{chp2:ddus:2}
	\mathcal{U}(x)=\left\{
	u=\mathcal{C}(u^{'},x)\ |\ u^{'}\in {U}^{\rm sub}
	\right\}
\end{gather}
\end{lemma}
The proof of Lemma \ref{lemma:1} is provided in the appendix. 

Based on Lemma \ref{lemma:1}, the ubiquity of complete separability of a DDUS can be justified by the following corollary.
\begin{corollary}
\label{corollary:1}
For any DDUS $\mathcal{U}(x):X\rightrightarrows U$, it is with complete separability as in \eqref{chp2:ddus:2} by setting the ${U}^{\rm sub}$ as $U$ and the coupling function $\mathcal{C}(\cdot)$ as:
\begin{subequations}
	\label{chp2:project}
	\begin{align}
		\mathcal{C}(u^{'},x)=\arg\quad \min_{u}\quad &\Vert u^{'}-u\Vert_2\\
		\text{s.t.}\quad & u\in \mathcal{U}(x)
	\end{align}
\end{subequations}
\end{corollary}
Corollary \ref{corollary:1} can be justified by checking that the coupling function in \eqref{chp2:project} satisfies the assumption in Lemma \ref{lemma:1}.

\subsubsection{Examples}
\begin{example}[]\label{example:1}
The adjustable uncertainty sets in \cite{zhang2017robust} are completely separable since they are restricted to having a specific family of geometric forms. Denote by $u$ the DDU variable and $\mathcal{U}(\cdot)$ its DDUS.
\begin{itemize}
	\item Ball DDUS: Denote by $r\ge 0$ the radius, $x$ the ball center, and $\Vert\cdot\Vert_p$ the $p$-norm of a vector. The ball DDUS $\mathcal{U}(x,r)$ parameterized in $x\in \mathbb{R}^n$ and $r\in \mathbb{R}^1$ is completely separable:
	\begin{align}
		\mathcal{U}(x,r)&:=\left\{u\ |\ \Vert u-x\Vert_p\le r\right\}\\
		&=\left\{u=\mathcal{C}(u^{'},x,r)\ |\ u^{'}\in\mathcal{U}^0\right\}
	\end{align}
where $u^{'}$ is the auxiliary uncertain variable with support set $\mathcal{U}^0:=\left\{u^{'}\ |\ \Vert u^{'} \Vert_p\le 1\right\}$ and the coupling function $\mathcal{C}(\cdot)$ is defined as $u=\mathcal{C}(u^{'},x,r)=x+rIu^{'}$ where $I\in\mathbb{R}^{n\times n}$ is an identity matrix.

\item Box DDUS: Denote by $x\in \mathbb{R}^n$ the center of the box and $r\in \mathbb{R}^n$ the range of the box. The box DDUS $\mathcal{U}(x,r)$ parameterized in $x\in\mathbb{R}^n$ and $r\in \mathbb{R}^n$ is completely separable:
\begin{subequations}
	\begin{align}
		\mathcal{U}(x,r)&:=\left\{u\ |\ \Vert u-x\Vert_{\infty}\le r\right\}\\
		&=\left\{u=\mathcal{C}(u^{'},x,r)\ |\ u^{'}\in\mathcal{U}^0\right\}
	\end{align}
\end{subequations}
where $u^{'}$ is the auxiliary uncertain variable with support set $\mathcal{U}^0:=\left\{u^{'}\ |\ \Vert u^{'} \Vert_{\infty}\le 1\right\}$ and the coupling function $\mathcal{C}(\cdot)$ is defined as $u=\mathcal{C}(u^{'},x,r)=x+\text{diag}(r)u^{'}$ where $\text{diag}(r)\in \mathbb{R}^{n\times n}$ is a diagonal matrix with diagonal elements $r$.

\item Ellipsoidal DDUS: Denote by $x\in\mathbb{R}^n$ the center of the ellipsoid and $\Sigma\in\mathbb{R}^{n\times n}$ a symmetric and positive definite matrix. The ellipsoidal DDUS $\mathcal{U}(x,\Sigma)$ parameterized in $x$ and $\Sigma$ is completely separable:
\begin{subequations}
	\begin{align}
		\mathcal{U}(x,\Sigma)&:=\left\{u\ |\ 
		(u-x)^{\mathsf{T}}\Sigma^{-1}(u-x)\le 1
		\right\}\\
		&=\left\{u=\mathcal{C}(u^{'},x,\Sigma)\ |\ u^{'}\in\mathcal{U}^0\right\}
	\end{align}
\end{subequations}
where $u^{'}$ is the auxiliary uncertain variable with support set $\mathcal{U}^0:=\left\{u^{'}\ |\ \Vert u^{'} \Vert_{2}\le 1\right\}$ and the coupling function $\mathcal{C}(\cdot)$ is defined as $u=\mathcal{C}(u^{'},x,\Sigma)=x+\Sigma^{1/2}u^{'}$.

\item Polyhedral DDUS: $\mathcal{U}(x^{(1)},\ldots,x^{(m)})$ parameterized in $x^{(1)},\ldots,x^{(m)}\in \mathbb{R}^n$ is the convex hall of $m$ ($m\ge n$) vectors $x^{(1)},\ldots,x^{(m)}$. Then, polyhedral DDUS $\mathcal{U}(x^{(1)},\ldots,x^{(m)})$ is completely separable as follows:
\begin{subequations}
	\begin{gather}
		\mathcal{U}(x^{(1)},\ldots,x^{(m)}):=\text{conv}(x^{(1)},\ldots,x^{(m)})\\
		\qquad\qquad=\left\{u=\mathcal{C}(u^{'},x^{(1)},\ldots,x^{(m)})\ |\ u^{'}\in\mathcal{U}^0\right\}
	\end{gather}
\end{subequations}
where $u^{'}$ is the auxiliary uncertain variable with support set $\mathcal{U}^0:=\text{conv}(e_1,\ldots,e_m)$ where $e_i\in\mathbb{R}^n$ is the unit vector with the $i$-th element being 1. The coupling function $\mathcal{C}(\cdot)$ is defined as $u=\mathcal{C}(u^{'},x^{(1)},\ldots,x^{(m)})=[x^{(1)},\ldots,x^{(m)}]u^{'}$.
\end{itemize}
\end{example}

\begin{example}
\label{example:2}
Denote by $P^{\rm mppt}$ the maximal available power of the wind turbine under the MPP tracking mode, $P^{\rm avail}$ the available power, $R$ the preserved power, and $v$ the uncertain wind speed. With given decisions on the loading scheme $I\in \{\text{\textit{De-rating}, \textit{Delta}, \textit{Percentage}}\}$ and the de-loading ratio $\lambda\in [0,1]$, the relationships between $P^{\rm avail}$, $R$ and $P^{\rm mppt}$ can be captured as: 
\begin{subequations}
	\label{example:2:wind-deload}
	\begin{align}
		\text{\textit{De-rating:}}\quad &\left\{
		\begin{array}{l}
			R=\max\left\{P^{\rm mppt}-(1-\lambda)P^{\rm rate},0\right\}\\
			P^{\rm avail}=\min\left\{
			P^{\rm mppt},(1-\lambda)P^{\rm rate}
			\right\}
		\end{array}
		\right.\\
		\text{\textit{Delta:}}\quad &\left\{
		\begin{array}{l}
			R=\min\left\{P^{\rm mppt},\lambda P^{\rm rate}\right\}\\
			P^{\rm avail}=\max\left\{
			P^{\rm mppt}-\lambda P^{\rm rate},0
			\right\}
		\end{array}
		\right. \label{example:2:wind-deload-delta} \\   
		\text{\textit{Percentage:}}\quad &\left\{
		\begin{array}{l}
			R=\lambda P^{\rm mppt}\\
			P^{\rm avail}=(1-\lambda) P^{\rm mppt}
		\end{array}
		\right.		
	\end{align}
\end{subequations}
where $P^{\rm mppt}$ is a DIU variable solely determined by the uncertain wind speed $v$:
\begin{gather}
	\label{example:2:mppt}
P^{\rm mppt}=\left\{
\begin{array}{ll}
	0,&v<v_{\rm in}\\
	\mu_0 v^3,&v_{\rm in}\le v\le v_{\rm rate}\\
	P^{\rm rate},&v_{\rm rate}\le v\le v_{\rm out}\\
	0,&v> v_{\rm out}
\end{array}
\right.
\end{gather}
Equations \eqref{example:2:wind-deload}-\eqref{example:2:mppt} can be viewed as the function of $P^{\rm avail}$ and $R$ with arguments $I$, $\lambda$, and $v$. Denote the compact form of this function by
\begin{gather}
\label{example:2:coupling}
\left[
\begin{array}{c}
	P^{\rm avail}\\
	R
\end{array}
\right] = \mathcal{C}(I,\lambda,v)
\end{gather}
Among the arguments of coupling function $\mathcal{C}(\cdot)$, $I$ and $\lambda$ are decision variables of the operator and the wind speed $v$ is an exogenously stochastic variate. Then, the DDUS of DDU variables $P^{\rm avail}$ and $R$ is completely separable with the aid of the coupling function $\mathcal{C}(\cdot)$ in \eqref{example:2:coupling} and the uncertainty set of $v$ denoted by $\mathcal{V}$:
\begin{gather}
\mathcal{U}(\lambda,I)=\left\{
\left[
\begin{array}{c}
	P^{\rm avail}\\
	R
\end{array}
\right]
=\mathcal{C}(I,\lambda,v)\ |\ v\in\mathcal{V}
\right\}
\end{gather}
\end{example}

\begin{example}
\label{example:dr:sep}
In Fig. \ref{fig:chp2:dr:data:two}, the DDUS $\mathcal{U}(x)$ is with complete separability as follows:
\begin{gather}
\label{example:dr:a:sep}
\mathcal{U}(x)=\left\{
u=g^{\rm lb}(x)+u^{'}\left(
g^{\rm ub}(x)-g^{\rm lb}(x)\right)\ |\ 
u^{'}\in [0,1]
\right\}
\end{gather}
In the separable formulation \eqref{example:dr:a:sep}, $u^{'}$ is the auxiliary random variable falling into range $[0,1]$ which is independent of decision variable $x$.

The DDUS of the uncertain demand $d$ in Fig.\ref{fig:dr:b} is also completely separable with formulation as follows:
\begin{gather}
\label{example:dr:b:sep}
\mathcal{U}(p)=\left\{
d=A(p+\epsilon)^{\alpha}\ |\ -\hat{\epsilon}\le \epsilon\le \hat{\epsilon}
\right\}
\end{gather}
The auxiliary random variable $\epsilon$ represents the deviation on the deterministic price-elastic demand curve and $\epsilon$ varies within the range $[-\hat{\epsilon},\hat{\epsilon}]$ which is irrelevant to decision $p$. 

In Fig.\ref{fig:dr:c}, by regarding the uncertain price-elastic coefficient $\epsilon$ as the decision-independent auxiliary random variable, then the DDUS of demand $d$ is completely separable:
\begin{gather}
\mathcal{U}(p)=\left\{
u=u_0\left(
1+\epsilon\frac{p-p_0}{p_0}
\right)\ |\ \epsilon\in [\epsilon^{\rm min},\epsilon^{\rm max}]
\right\}
\end{gather}
\end{example}

\subsection{The Partial Separability of DDUS}
\label{subsec:partial_separa}

\subsubsection{Definition}
We define the partial separability of a DDUS as follows.
\begin{definition}[Partial Separability of DDUS]
\label{def:sep:partial}	
If the following two conditions hold for a DDUS $\mathcal{U}(x)$, then $\mathcal{U}(x)$ is called \textbf{partially separable}.
\begin{itemize}
	\item [a)] There exists a set-valued map $\mathcal{U}^{\rm sub}(x):X\rightrightarrows U$ such that the graph of $\mathcal{U}^{\rm sub}(x)$ is the subset of the graph of $\mathcal{U}(x)$, i.e., 
	$$
	\text{graph}(\mathcal{U}^{\rm sub}(\cdot))\subseteq \text{graph}(\mathcal{U}(\cdot)).
	$$
	\item [b)] The set-valued map $\mathcal{U}^{\rm sub}(x)$ in condition a) is separable with the following formulation:
	$$
	\mathcal{U}^{\rm sub}(x)=\left\{
	u=\mathcal{C}(\xi,x)\ |\ \xi\in \Xi
	\right\},\forall x\in X.
	$$
\end{itemize}
\end{definition}

\subsubsection{Ubiquity} 
Considering the DDU-integrated TSRO problem \eqref{tsro}, we would like to show that any given DDUS $\mathcal{U}(x)$ is partially separable and can be equivalently substituted by the set-valued map with complete separability. 

For the TSRO problem \eqref{tsro} and any given DDUS $\mathcal{U}(x)$, define the separable DDUSs with regard to robust feasibility and robust optimality respectively as follows:
\begin{itemize}
	\item [a)] The DDUS $\mathcal{U}^{\rm sep}_{\rm fea}(x):X\rightrightarrows U$ is formulated considering the robust feasibility of TSRO problem \eqref{tsro} as follows:
	\begin{subequations}
		\label{sep:fea}
		\begin{align}
			\label{sep:fea:1}
			\mathcal{U}_{\rm fea}^{\rm sep}(x):=\left\{
			u=\mathcal{C}_{\rm fea}(\xi,x)\ |\ \xi\in \Xi
			\right\},\forall x\in X
		\end{align}
		where $\Xi$ is the decision-independent support of auxiliary random variable $\xi$:
		\begin{align} 
			\label{sep:fea:2}
			\Xi:=\left\{
			\xi\ |\ B^{\mathsf{T}}\xi \le \mathbf{0},-\mathbf{1}\le \xi\le \mathbf{0}
			\right\}
		\end{align}
		and $\mathcal{C}_{\rm fea}(\cdot)$ is the coupling function defined as follows:
		\begin{align}
			\label{sep:fea:3}
			\mathcal{C}_{\rm fea}(\xi,x) := \arg \left\{
			\begin{array}{ll}
				\max_{u}\ &-\xi^{\mathsf{T}}Cu\\
				\text{s.t.}\ &u\in\mathcal{U}(x)
			\end{array}
			\right\}
		\end{align}
	\end{subequations}
	
	\item [b)] The DDUS $\mathcal{U}^{\rm sep}_{\rm opt}(x):X\rightrightarrows U$ is formulated considering the robust optimality of TSRO problem \eqref{tsro} as follows:
	\begin{subequations}
		\label{sep:opt}
		\begin{align}
			\label{sep:opt:1}
			\mathcal{U}_{\rm opt}^{\rm sep}(x):=\left\{
			u=\mathcal{C}_{\rm opt}(\pi,x)\ |\ \pi\in \Pi
			\right\},\forall x\in X
		\end{align}
		where $\Pi$ is the decision-independent support of auxiliary random variable $\pi$:
	    \begin{align}
			\label{sep:opt:2}
			\Pi:=\left\{
			\pi\ |\ B^{\mathsf{T}}\pi \le c,\pi\le \mathbf{0}
			\right\}
		\end{align}
		and $\mathcal{C}_{\rm opt}(\cdot)$ is the coupling function formulated as follows:
		\begin{align}
			\label{sep:opt:3}
			\mathcal{C}_{\rm opt}(\pi,x) := \arg \left\{
			\begin{array}{ll}
				\max_{u}\ &-\pi^{\mathsf{T}}Cu\\
				\text{s.t.}\ &u\in\mathcal{U}(x)
			\end{array}
			\right\}
		\end{align}
	\end{subequations}
\end{itemize}

First of all, it is easy to verify that the graph of $\mathcal{U}_{\rm fea}^{\rm sep}(x)$ and $\mathcal{U}_{\rm opt}^{\rm sep}(x)$ are the subsets of the graph of $\mathcal{U}(x)$, therefore the condition a) in Definition \ref{def:sep:partial} is satisfied. Combining with the fact that  $\mathcal{U}_{\rm fea}^{\rm sep}(x)$ and $\mathcal{U}_{\rm opt}^{\rm sep}(x)$ are completely separable, it is revealed that the DDUS $\mathcal{U}(x)$ is with partial separability regardless of the specific formulation of $\mathcal{U}(x)$.

Next, we would like to justify that the DDUS $\mathcal{U}(x)$ in TSRO problem \eqref{tsro} can be equivalently substituted by the two separable DDUSs $\mathcal{U}_{\rm fea}^{\rm sep}(x)$ and $\mathcal{U}_{\rm opt}^{\rm sep}(x)$.
\begin{theorem}
	\label{thm:eq}
	The TSRO problem \eqref{tsro} with arbitrary DDUS $\mathcal{U}(x)$ has surrogate model as follows:
\begin{itemize}
	\item [a)] The RFR $X_R$ is with the following equivalent formulation:
	\begin{subequations}
		\label{xr:eq}
		\begin{align}
			\label{xr:eq:1}
			X_R&\overset{(i)}{:=}\left\{x\in X\ |\ Y(x,u)\neq \emptyset,\forall u\in \mathcal{U}(x)\right\}\\
			\label{xr:eq:2}
			&\overset{(ii)}{=}\left\{x\in X\ |\ Y(x,u)\neq \emptyset,\forall u\in \mathcal{U}_{\rm fea}^{\rm sep}(x)\right\}
		\end{align}
	\end{subequations}
	where $(i)$ is according to the definition of $X_R$ in \eqref{tsro:4} and $(ii)$ implies the surrogate formulation of $X_R$.
	\item [b)] The robust optimality sub-function $S(x)$ is with the following equivalent formulation:
	\begin{subequations}
	\label{s:eq}
	\begin{align}
		S(x)\overset{(i)}{:=}&\max_{u\in \mathcal{U}(x)}\min_{y\in Y(x,u)}\ c^{\mathsf{T}}y,\quad\forall x\in X\\
		\overset{(ii)}{=}&\max_{u\in \mathcal{U}_{\rm opt}^{\rm sep}(x)}\min_{y\in Y(x,u)}\ c^{\mathsf{T}}y,\quad\forall x\in X
	\end{align}
	\end{subequations}
	where $(i)$ is according to the definition of $S(x)$ in \eqref{tsro:3} and $(ii)$ implies the surrogate formulation of function $S(x)$.
\end{itemize}
\end{theorem}
The proof of Theorem \ref{thm:eq} is provided in the appendix.

According to Theorem \ref{thm:eq}, for any DDUS $\mathcal{U}(x)$, the TSRO problem \eqref{tsro} is equivalent with \eqref{tsro-eq} as follows
\begin{subequations}
	\label{tsro-eq}
\begin{gather}
\min_{x}\ 
\left\{f(x)+\max_{u\in \mathcal{U}_{\rm opt}^{\rm sep}(x)}\min_{y\in Y(x,u)}\ c^{\mathsf{T}}y\right\}\\
\text{s.t.}\quad x\in X\\
Y(x,u)\neq \emptyset,\forall u\in \mathcal{U}_{\rm fea}^{\rm sep}(x)
\end{gather}
\end{subequations}
where $\mathcal{U}_{\rm opt}^{\rm sep}(x)$ and $\mathcal{U}_{\rm fea}^{\rm sep}(x)$ are two separable DDUSs. 

\subsubsection{Example}
\label{example:8}
Next, an illustration example is provided to show how the partial separability of DDUS is applied to the TSRO problem. 

\begin{example}
\label{example8}
Consider the polyhedral DDUS as follows:
	\begin{gather}
	\label{chp2:eg:poly}
	\mathcal{U}(x)=\left\{
	u\in\mathbb{R}^2\ |
	\begin{array}{c}
		u_1\le 7x_1+8x_2\\
		u_2\le 13 x_2\\
		-u_1+2u_2\le 15x_2+8\\
		u_1+u_2\le 7x_1+2x_2+13\\
		4u_1-7u_2\le 21 x_1+11x_2-25\\
		-8u_1-3u_2\le -40
	\end{array} 
	\right\}
\end{gather}
where the here-and-now decision variable $x=(x_1,x_2)^{\mathsf{T}}\in\mathbb{R}^2$ determines the right-hand-side vector of the polytope uncertainty set. 

It is shown in Fig. \ref{fig:ddus:poly} that, the shape, size, and location of the DDUS $\mathcal{U}(x)$ in \eqref{chp2:eg:poly} changes with the different values of the decision variable $x$. 

For a TSRO problem with $Y(x,u)$ in \eqref{chp2:example:II:Y},
\begin{gather}
	\label{chp2:example:II:Y}
	Y(x,u)=\left\{
	y\in\mathbb{R}^2\ |\ \begin{array}{c}
		-1\le y_1\le 1\quad \left(\xi_1,\xi_2\right)\\
		-1\le y_2\le 1\quad \left(\xi_3,\xi_4\right)\\
		u_1\le y_1+y_2\le u_2\quad \left(\xi_5,\xi_6\right)
	\end{array} 
	\right\}
\end{gather}
an auxiliary DDUS $\mathcal{U}_{\rm fea}^{\rm sep}(x)$ concerning robust feasibility is formulated as in \eqref{eg:fea:sep}.
\begin{subequations}
\label{eg:fea:sep}
\begin{gather}
	\label{chp2:example:II:U:SEP}
	\mathcal{U}_{\rm fea}^{\rm sep}(x):=\left\{
	u=\mathcal{C}_{\rm fea}(\xi,x)\ |\ \xi\in \Xi
	\right\},\forall x\in X
\end{gather}
\begin{gather}
	\label{chp2:example:II:Xi}
	\Xi:=\left\{
	\xi\in\mathbb{R}^6\ |\ 
	\begin{array}{c}
		\xi_5-\xi_6=\xi_2-\xi_1\\
		\xi_6-\xi_5=\xi_3-\xi_4\\
		-1\le \xi_i\le 0,\forall i=1,\ldots,6
	\end{array}
	\right\}
\end{gather}
\begin{gather}
	\label{chp2:example:II:C}
	\mathcal{C}_{\rm fea}(\xi,x):=\arg\left\{
	\begin{array}{ll}
		\max_{u}\ &-\xi_5 u_1+\xi_6 u_2\\
		\text{s.t.}\ &u\in \mathcal{U}(x)
	\end{array}
	\right\}
\end{gather}
\end{subequations}

Both $\mathcal{U}(x)$ and $\mathcal{U}_{\rm fea}^{\rm sep}(x)$ vary with the decision variable $x$. It can be verified that $\mathcal{U}(x)$ and $\mathcal{U}_{\rm fea}^{\rm sep}(x)$ satisfy the conditions in Definition \ref{def:sep:partial}. Therefore, $\mathcal{U}(x)$ can be partially separated and $\mathcal{U}_{\rm fea}^{\rm sep}(x)$ is completely separable.

\begin{figure*}[!htb]
	\centering
	\subfloat[$x=(1,1)^{\mathsf{T}}$]{\label{fig:ddus:poly:a}
		\includegraphics[width=0.3\linewidth]{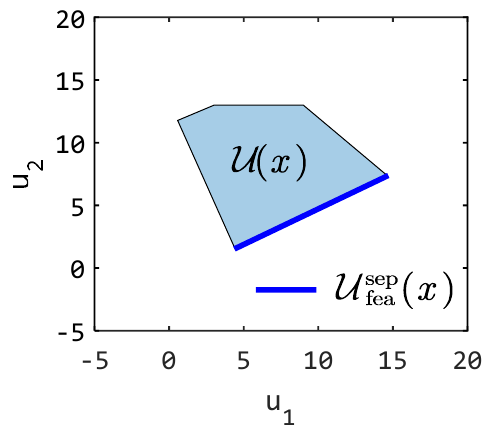}}
	\subfloat[$x=(0,1)^{\mathsf{T}}$]{\label{fig:ddus:poly:b}
		\includegraphics[width=0.3\linewidth]{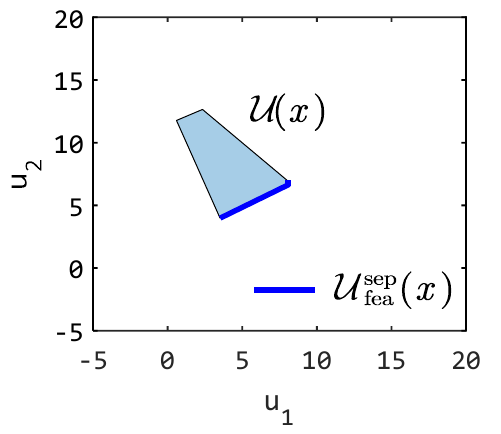}} 
	\subfloat[$x=(0.5,0.5)^{\mathsf{T}}$]{\label{fig:ddus:poly:c}
		\includegraphics[width=0.3\linewidth]{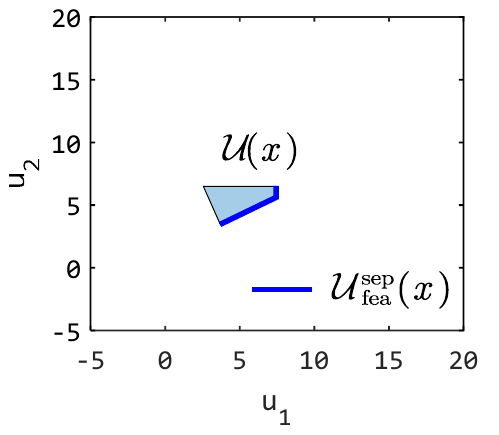}}
	\caption{Polyhedral decision-dependent uncertainty set and its partial separability.}
	\label{fig:ddus:poly} 
\end{figure*}

As illustrated in Fig. \ref{fig:ddus:poly}, for any given $x\in\mathbb{R}^2$, $\mathcal{U}_{\rm fea}^{\rm sep}(x)$ is the subset of the original DDUS $\mathcal{U}(x)$. According to Theorem \ref{thm:eq}, the RFR $X_R$ regarding $x$ has two surrogate formulations, one based on $\mathcal{U}(x)$ in \eqref{xr:eq:1} and the other based on $\mathcal{U}_{\rm fea}^{\rm sep}(x)$ in \eqref{xr:eq:2}.
\end{example}

\section{Mechanism of Robust Dispatch with Decision-Dependent Uncertainties}
In this section, we first derive the surrogate model of problem \eqref{tsro} based on the concept of the dispatchable region \cite{wei2015dispatchable,wang2016risk} and its variants. Then, the mechanism of TSRO dispatch with DDUs is revealed.

\begin{definition}[Dispatchable Region]
\label{def:disp-reg}
The dispatchable region of problem \eqref{tsro}, denoted by $\mathcal{D}(x):X\rightrightarrows U$, is defined by \eqref{def-dis-reg}.
\begin{subequations}
\label{def-dis-reg}
\begin{align}
\forall u\in \mathcal{D}(x),\ Y(x,u)\neq \emptyset\\
\forall u\notin \mathcal{D}(x),\ Y(x,u)= \emptyset
\end{align}
\end{subequations}

The extended dispatchable region of problem \eqref{tsro}, denoted by $\mathcal{D}^{\rm ext}(x,\alpha):X\times\mathbb{R}^1\rightrightarrows U$ with arguments $x\in X$ and $\alpha\in\mathbb{R}^1$, is defined by \eqref{def-ex-dis-reg}.
\begin{subequations}
	\label{def-ex-dis-reg}
	\begin{align}
		\forall u\in \mathcal{D}^{\rm ext}(x,\alpha),\ Y(x,u)\cap \left\{
		y|c^{\mathsf{T}}y\le \alpha
		\right\}\neq \emptyset\\
		\forall u\notin \mathcal{D}^{\rm ext}(x,\alpha),\ Y(x,u)\cap \left\{
		y|c^{\mathsf{T}}y\le \alpha
		\right\}= \emptyset
	\end{align}
\end{subequations}
It can be easily verified that for any $x\in X$ and $\alpha\in\mathbb{R}^1$, $\mathcal{D}^{\rm ext}(x,\alpha)\subseteq \mathcal{D}(x)$.
\end{definition}

According to the definition of the extended dispatchable region, the surrogate model of \eqref{tsro} is established as \eqref{eq-tsro}.
\begin{subequations}
\label{eq-tsro}
\begin{gather}
\label{eq-tsro:1}
\min_{x\in X,\alpha\in \mathbb{R}^1}\ f(x)+\alpha\\
\label{eq-tsro:2}
\text{s.t.}\ \mathcal{U}(x)\subseteq \mathcal{D}^{\rm ext}(x,\alpha)
\end{gather}
\end{subequations}

In \eqref{eq-tsro}, $\alpha$ is an auxiliary decision variable for denoting the value of function $S(x)$. Constraint \eqref{eq-tsro:2} combines the robust feasibility requirement $x\in X_R$ and the robust optimality function $S(x)$. Note that the equivalence between \eqref{tsro} and \eqref{eq-tsro} has no relation with the specific formulation of DDUS $\mathcal{U}(x)$. Therefore, this equivalence also applies to DIU sets as DIU is a special case of DDU.

Based on problem \eqref{eq-tsro}, the mechanism of TSRO hedging against uncertainties can be concluded as the optimal match between the uncertainty set $\mathcal{U}(x)$ and the extended dispatchable region $\mathcal{D}^{\rm ext}(x,\alpha)$. Constraint \eqref{eq-tsro:2} enforces that the extended dispatch region $\mathcal{D}^{\rm ext}(x,\alpha)$ must cover the uncertainty set $\mathcal{U}(x)$. The objective \eqref{eq-tsro:1} aims at minimizing the matching cost which is $f(x)+\alpha$.

\begin{figure*}[!htb]
	\centering
	\subfloat[Robust dispatch under DIUs.]{\label{fig:mechanism:diu}
		\includegraphics[width=0.45\linewidth]{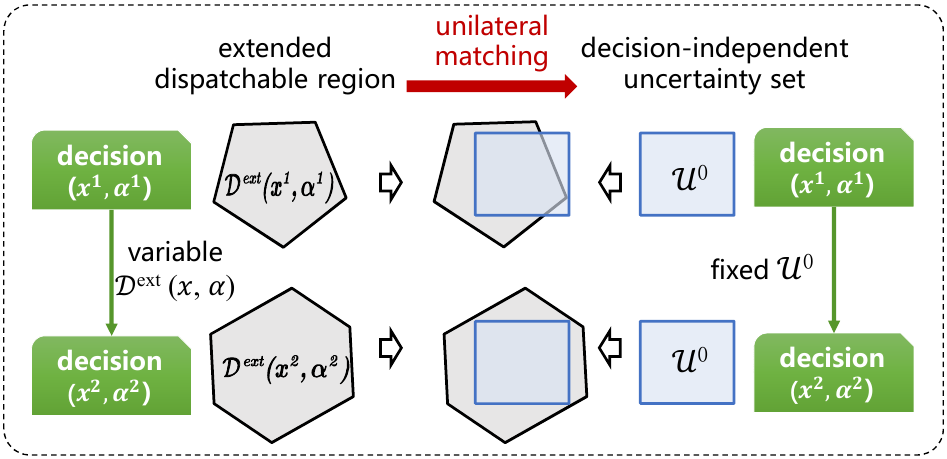}}
	\subfloat[Robust dispatch under DDUs.]{\label{fig:mechanism:ddu}
		\includegraphics[width=0.45\linewidth]{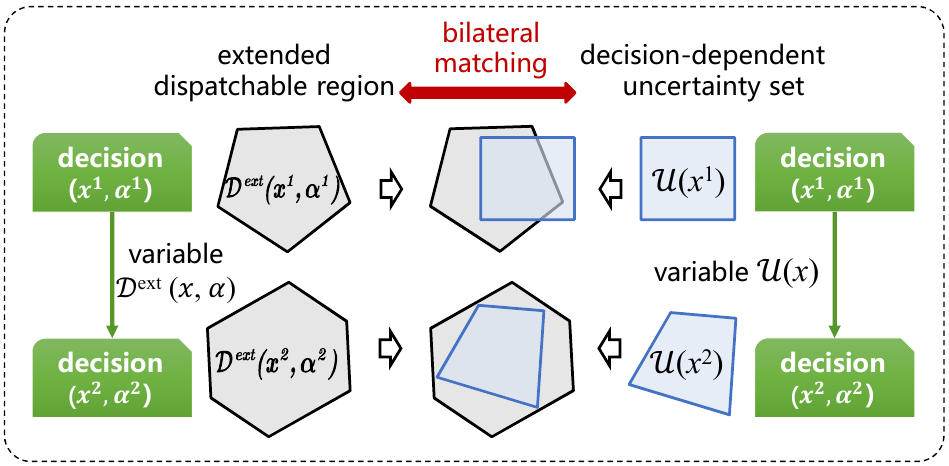}} 
	\caption{The mechanism of robust dispatch under DIUs and DDUs.}
	\label{fig:mechanism}
\end{figure*}

The difference in the mechanism of robust dispatch under DIUs and DDUs is illustrated in Fig. \ref{fig:mechanism}. In the case of DIU, the decision maker has to find a feasible $(x,\alpha)$ such that the extended dispatchable region $\mathcal{D}^{\rm ext}(x,\alpha)$ that varies with $(x,\alpha)$ covers the fixed uncertainty set $\mathcal{U}^0$. This process can be concluded as the unilateral matching, where the dispatchable region is adapted to suit the uncertainty set. In the case of DDU, both the extended dispatchable region $\mathcal{D}^{\rm ext}(x,\alpha)$ and the DDUS $\mathcal{U}(x)$ vary with decision variables $(x,\alpha)$ and mechanism of robust dispatch under DDUs can be summarized as the bilateral matching between they two.

\begin{remark}[Bilateral Matching]
The bilateral matching mechanism points out the key difference between DDU and DIU in power system robust dispatch. Instead of a burden, DDU is actually an active flexible source. With rational decisions, the DDUS can be reshaped so that the uncertainty can be more easily admitted. In other words, the potential flexibility in DDU is explored by the bilateral matching mechanism, which would inspire the dispatch and operation in RES-dominated power systems.
\end{remark}

According to the subsections \ref{subsec:complete_separa}
and \ref{subsec:partial_separa}, there exist both the complete and partial separable formulations for the DDUS $\mathcal{U}(x)$ in problem \eqref{eq-tsro}. Suppose that $\mathcal{U}(x)$ is equivalent to or can be equivalently substituted by a separable DDUS $\mathcal{U}^{\rm sep}(x)=\{
u=\mathcal{C}(\xi,x)|\xi\in \Xi
\}$. Then, it can be verified that constraint \eqref{eq-tsro:2} can be transformed into:
\begin{gather}
\eqref{eq-tsro:2}\iff \Xi\subseteq 
\mathfrak{D}(x,\alpha)
\end{gather}
where $\mathfrak{D}(x,\alpha)$ can be viewed as the extended dispatchable region of the auxiliary random variable $\xi$:
\begin{gather}
\label{def:mathfrak:D}
\mathfrak{D}(x,\alpha):=\left\{
\xi|\mathcal{C}(x,\xi)\in \mathcal{D}^{\rm ext}(x,\alpha)
\right\}
\end{gather}
Therefore, the bilateral matching between the extended dispatchable region $\mathcal{D}^{\rm ext}(x,\alpha)$ and the DDUS $\mathcal{U}(x)$ can be reduced to the unilateral from $\mathfrak{D}(x,\alpha)$ to the fixed $\Xi$. An example is provided as follows.

\begin{example}
\label{example9}
On the basis of Example \ref{example8}, we leave the robust optimality and see how the dispatchable region and the uncertainty set match with each other. 

For DDU variable $u$ with DDUS $\mathcal{U}(x)$ in \eqref{chp2:eg:poly}, it is straightforward to obtain the dispatchable region of $u$:
\begin{gather}
\label{disp-reg}
\mathcal{D}(x)=\left\{
u\in\mathbb{R}^2|
u_1\le 2,u_2\ge -2, u_1\le u_2
\right\}
\end{gather}

According to Example \ref{example8}, $\mathcal{U}(x)$ can be equivalently substituted by the separable DDUS 
$\mathcal{U}_{\rm fea}^{\rm sep}(x)$ in \eqref{eg:fea:sep}, where there exists an auxiliary random variable $\xi$ with decision-independent support $\Xi$ in \eqref{chp2:example:II:Xi}. The dispatchable region of $\xi$, denoted by $\mathfrak{D}(x)$, can be obtained according to \eqref{def:mathfrak:D}. Next, we would like to show that
\begin{subequations}
\begin{align}
x\in X_R&\iff \mathcal{U}(x)\subseteq \mathcal{D}(x)\\
&\iff \mathcal{U}_{\rm fea}^{\rm sep}(x)\subseteq \mathcal{D}(x)\\
&\iff \Xi\subseteq \mathfrak{D}(x)
\end{align}
\end{subequations}

When $x=(1,1)^{\mathsf{T}}$, the corresponding $\mathcal{D}(x)$, $\mathcal{U}(x)$, and $\mathcal{U}_{\rm fea}^{\rm sep}(x)$ are plotted in Fig. \ref{fig:example9:DDUS1} (a); the corresponding $\mathfrak{D}(x)$ and $\Xi$ are plotted in Fig. \ref{fig:example9:DDUS1} (b). It is shown that $\mathcal{D}(x)$ is not able to cover neither $\mathcal{U}(x)$ nor $\mathcal{U}_{\rm fea}^{\rm sep}(x)$. Moreover, $\mathfrak{D}(x)$ does not cover $\Xi$. These indicate that $x=(1,1)^{\mathsf{T}}$ is not robust feasible.

\begin{figure}[!htb]
	\centering
	\subfloat[The space of $u$.]{\label{fig:example9:DDUS1:a}
		\includegraphics[width=0.43\linewidth]{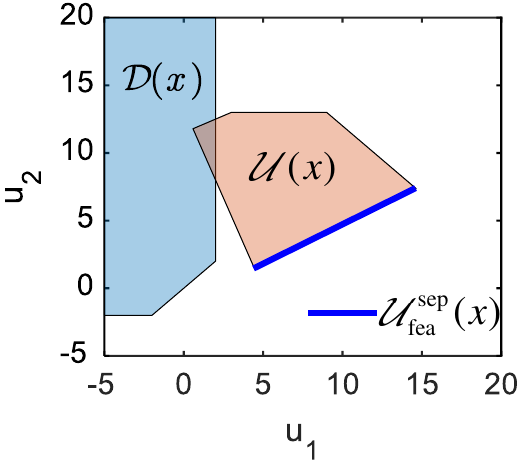}}
	\subfloat[The space of $\xi$.]{\label{fig:example9:DDUS1:b}
		\includegraphics[width=0.43\linewidth]{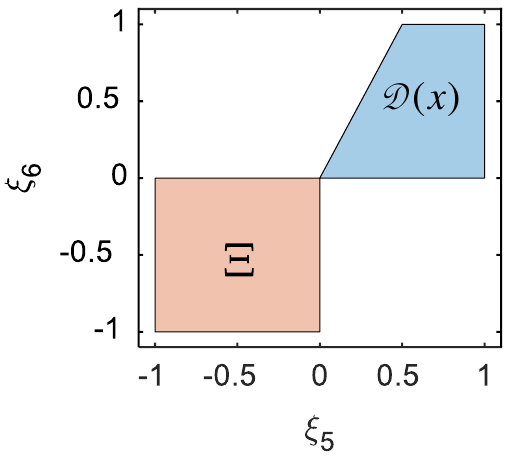}} 
	\caption{When fixing $x=(1,1)^{\mathsf{T}}$, the DDUS \eqref{chp2:eg:poly}, the dispatchable region of $u$ denoted by $\mathcal{D}(x)$, the uncertainty set of $\xi$ denoted by $\Xi$, and the dispatchable region of $\xi$ denoted by $\mathfrak{D}(x)$.}
	\label{fig:example9:DDUS1} 
\end{figure}

With other conditions unchanged, when the DDUS $\mathcal{U}(x)$ is modified into:
\begin{gather}
	\label{chp2:eg:poly:2}
	\mathcal{U}(x)=\left\{
	u\in\mathbb{R}^2\ |
	\begin{array}{c}
		u_1\le x_1+x_2\\
		u_2\le 13 x_2\\
		-u_1+2u_2\le 15x_2+8\\
		u_1+u_2\le 7x_1+2x_2+13\\
		4u_1-7u_2\le 21 x_1+11x_2-25\\
		-8u_1-3u_2\le -40
	\end{array} 
	\right\}
\end{gather}
The dispatchable region of $u$ and the uncertainty set of $\xi$ remain unchanged whereas $\mathcal{U}_{\rm fea}^{\rm sep}(x)$ and $\mathfrak{D}(x)$ vary with $\mathcal{U}(x)$. By fixing $x=(1,1)^{\mathsf{T}}$, these regions are plotted in Fig. \ref{fig:example9:DDUS2}. It is observed that, in the space of $u$, $\mathcal{D}(x)$ covers both $\mathcal{U}(x)$ and $\mathcal{U}_{\rm fea}^{\rm sep}(x)$. In the space of $\xi$, $\mathfrak{D}(x)$ covers $\Xi$. These indicate that $x=(1,1)^{\mathsf{T}}$ is robust feasible when the DDUS is modified into \eqref{chp2:eg:poly:2}.

\begin{figure}[!htb]
	\centering
	\subfloat[The space of $u$.]{\label{fig:example9:DDUS2:a}
		\includegraphics[width=0.43\linewidth]{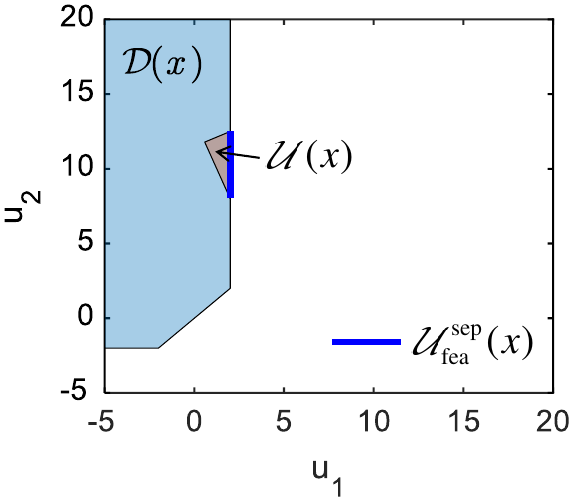}}
	\subfloat[The space of $\xi$.]{\label{fig:example9:DDUS2:b}
		\includegraphics[width=0.43\linewidth]{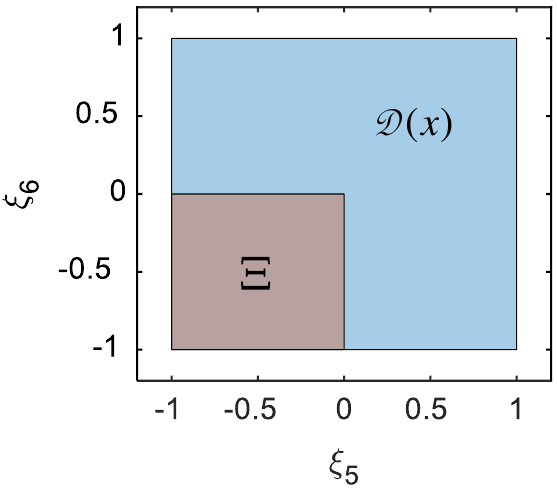}} 
	\caption{When fixing $x=(1,1)^{\mathsf{T}}$, the DDUS \eqref{chp2:eg:poly:2}, the dispatchable region of $u$ denoted by $\mathcal{D}(x)$, the uncertainty set of $\xi$ denoted by $\Xi$, and the dispatchable region of $\xi$ denoted by $\mathfrak{D}(x)$.}
	\label{fig:example9:DDUS2} 
\end{figure}

\end{example}

\section{Convexity of Robust Dispatch Problem with Decision-Dependent Uncertainties}
\label{sec:convexity}
DDUs facilitate the bilateral matching between the dispatchable region and the uncertainty set. However, compared with DIUs, the challenge of achieving the optimal matching is heightened. This section reveals that DDUs may introduce non-convexity to the TSRO problem.
\subsection{Case of DIUs}
It has been justified in \cite{wei2013phd} that for a TSRO problem that contains only linear constraints and DIUs, its RFR is convex. This conclusion can be extended to the convexity of the entire TSRO problem, see Lemma \ref{lemma:diu:convex} as follows.

\begin{lemma}
\label{lemma:diu:convex}
Let Assumption \ref{assp-tsro} hold for problem \eqref{tsro}. If $\mathcal{U}(x)=\mathcal{U}^0,\forall x\in X$ and $\mathcal{U}^0$ is a polyhedron, then the problem \eqref{eq-tsro} is convex.
\end{lemma}
The proof of Lemma \ref{lemma:diu:convex} is provided in the appendix \ref{app-C}. Note that Lemma \ref{lemma:diu:convex} only guarantees the convexity of TSRO problem \eqref{eq-tsro} when the uncertainty set is both decision-independent and polyhedral. Next, it would be revealed in Corollary \ref{coro:convex} that the convexity of \eqref{eq-tsro} actually relies on the decision-independence of the uncertainty set rather than its polyhedrality.

\subsection{Case of DDUs}
In this subsection, we would like to prove that when the uncertainty set is decision-dependent, the convexity of TSRO problem \eqref{eq-tsro} is no longer guaranteed. 

\subsubsection{DDUS with Complete Separability}
Without loss of generality, it is assumed that the DDUS $\mathcal{U}(x)$ is separable as 
\begin{gather}
	\label{thm:convex:sep}
\mathcal{U}(x)=\left\{
u=\mathcal{C}(\xi,x)|\xi\in \Xi
\right\}
\end{gather}
Then, Theorem \ref{thm:convex} gives the sufficient condition for the convexity of TSRO problem \eqref{eq-tsro}.
\begin{theorem}
\label{thm:convex}
Let Assumption \ref{assp-tsro} hold for problem \eqref{tsro}. If $\mathcal{U}(x)$ is separable as in \eqref{thm:convex:sep}, and the composite function $\mathcal{T}(\xi,x):\text{conv}(\xi)\times \mathbb{R}^{n_x}\rightarrow \mathbb{R}^m$ in \eqref{def:composite:function} is convex, then the TSRO problem \eqref{eq-tsro} is convex.
\begin{gather}
\label{def:composite:function}
\mathcal{T}(\xi,x):=C\times \mathcal{C}(\xi,x)
\end{gather}
In \eqref{def:composite:function}, $C$ is the matrix constant in \eqref{tsro:5}.
\end{theorem}
The proof of Theorem \ref{thm:convex} is provided in the appendix \ref{app-D}.

Theorem \ref{thm:convex} implies that, when the DDUS $\mathcal{U}(x)$ is separable, the convexity of the TSRO problem \eqref{eq-tsro} can be determined by the property of the coupling function $\mathcal{C}(\cdot)$ and has no other requirement on the decision-independent auxiliary uncertainty set $\Xi$, as long as $\Xi$ is independent of decision $x$.

A corollary of Theorem \ref{thm:convex} is provided as follows.
\begin{corollary}
\label{coro:convex}
Let Assumption \ref{assp-tsro} hold for problem \eqref{tsro}. If $\mathcal{U}(x)=\mathcal{U}^0,\forall x\in X$, then the TSRO problem \eqref{eq-tsro} is convex.
\end{corollary}

Corollary \ref{coro:convex} implies that when the uncertainty set is independent of the decision variable $x$, the convexity of TSRO problem \eqref{tsro} and \eqref{eq-tsro} holds and this has no relation with the formulation of the DIUS $\mathcal{U}^0$.

An example to illustrate Theorem \ref{thm:convex} is provided next.
\begin{example}
\label{example10}
On the basis of Example \ref{example8}, consider the DDUS $\mathcal{U}(x)$ as follows:
\begin{subequations}
\label{U-example10}
\begin{gather}
\mathcal{U}(x)=\left\{
u=\xi+x|\xi\in\Xi
\right\}\\
\Xi=\left\{
\mathbf{0}\le \xi\le \mathbf{1}
\right\}\cup\left\{
-\mathbf{1}\le \xi\le \mathbf{0}
\right\}
\end{gather}
\end{subequations}
where $x,u,\xi\in \mathbb{R}^2$. 

Since the $\mathcal{U}(x)$ in \eqref{U-example10} is separable and the coupling function is convex, according to Theorem \ref{thm:convex}, the corresponding RFR $X_R$ is convex, even though the DIUS $\Xi$ is nonconvex. To verify this, by searching all over the $\mathbb{R}^2$ space, the RFR $X_R$ that contains all robustly feasible $x$ is plotted in Fig. \ref{fig:example10}. The result complies with Theorem \ref{thm:convex}.
\begin{figure}[!htb]
	\centering
	\subfloat[$X_R$]{\label{fig:example10:a}
		\includegraphics[width=0.43\linewidth]{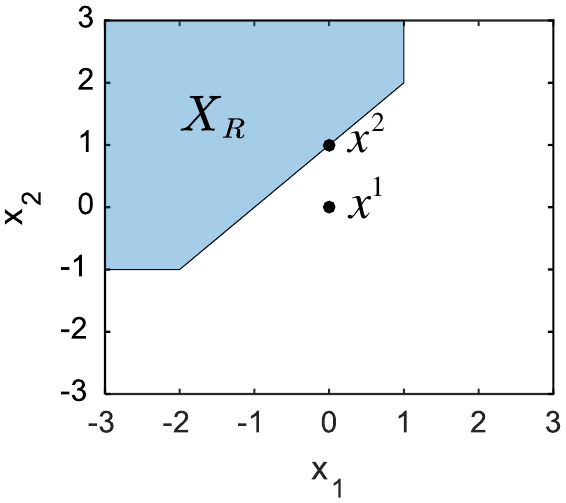}}
	\subfloat[$\Xi$]{\label{fig:example10:b}
		\includegraphics[width=0.43\linewidth]{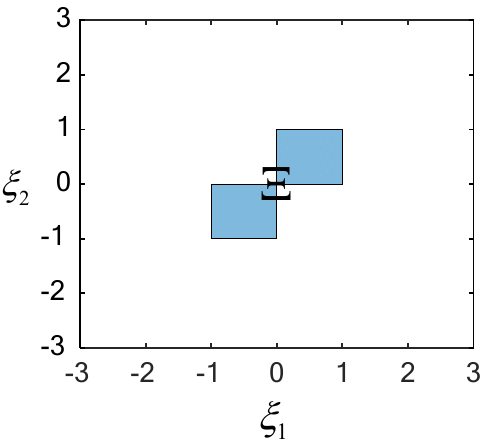}} 
	\caption{(a) The RFR $X_R$. (b) The DIUS $\Xi$.}
	\label{fig:example10} 
\end{figure}

Moreover, denote by $x^1=(0,0)^{\mathsf{T}}$ and $x^2=(0,1)^{\mathsf{T}}$. It is observed in Fig. \ref{fig:example10} (a) that, $x_1\notin X_R$ is not robustly feasible whereas $x^2\in X_R$ is. To verify this from another side, the dispatchable region $\mathcal{D}(x)$ and the DDUS $\mathcal{U}(x)$ with different values of $x$ are plotted in Fig. \ref{fig:example10-2}.
\begin{figure}[!htb]
	\centering
	\subfloat[$x=x^1$]{\label{fig:example10:c}
		\includegraphics[width=0.43\linewidth]{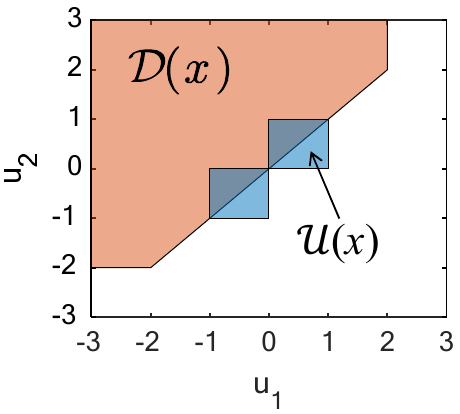}}
	\subfloat[$x=x^2$]{\label{fig:example10:d}
		\includegraphics[width=0.43\linewidth]{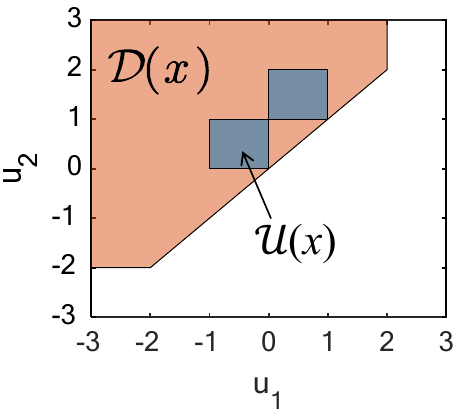}} 
	\caption{The dispatchable region $\mathcal{D}(x)$ and the DDUS $\mathcal{U}(x)$ when (a) $x=x^1$ and (b) $x=x^2$.}
	\label{fig:example10-2} 
\end{figure}

It is revealed in Fig. \ref{fig:example10-2} (a) that when $x=x^1$, the dispatchable region $\mathcal{D}(x^1)$ is not able to cover the DDUS $\mathcal{U}(x^1)$, complying with $x^1\notin X_R$. In Fig. \ref{fig:example10-2} (b), when $x=x^2$, the DDUS $\mathcal{U}(x^2)$ is covered by the dispatchable region $\mathcal{D}(x^2)$, which is compatible with $x^2\in X_R$.
\end{example}

\subsubsection{DDUS with Partial Separability}
According to Lemma \ref{lemma:diu:convex} and Theorem \ref{thm:convex}, the TSRO problem \eqref{eq-tsro} with decision-independent polyhedral uncertainty set is guaranteed to be convex. However, when the polytope varies with the here-and-now decision $x$, the TSRO problem \eqref{eq-tsro} could become nonconvex, see the example as follows.

\begin{example}
\label{example11}
On the basis of Example \ref{example8}, consider the DDUS $\mathcal{U}(x)$ as follows where the here-and-now decision $x\in\mathbb{R}^1$ varies within $[0.8,2.2]$.
\begin{gather}
	\label{chp2:eg:dduset}
	\mathcal{U}(x)=\left\{
	u\in\mathbb{R}^2\ |
	\begin{array}{c}
		u_1\le 6-2x\\
		u_1\le 2x\\
		u_2\le 13 x\\
		-u_1+2u_2\le 15x+8\\
		u_1+u_2\le 31-9x\\
		4u_1-7u_2\le 32x-25\\
		-8u_1-3u_2\le -20\\
		0\le u_1\le 3\\
		8\le u_2\le 13
	\end{array} 
	\right\}
\end{gather}
As the $Y(x,u)$ is defined in \eqref{chp2:example:II:Y}, the dispatchable region $\mathcal{D}(x)$ remains \eqref{disp-reg}. To obtain the RFR $X_R$, the graph of $\mathcal{U}(x)$ and $\mathcal{D}(x)$ are plotted in Fig. \ref{fig:example:11}. Moreover, $\mathcal{U}(x)$ and $\mathcal{D}(x)$ with different values of $x$ are shown in Fig. \ref{fig:example:11-2}.

\begin{figure}[!htb]
	\centering
	\includegraphics[width=0.8\linewidth]{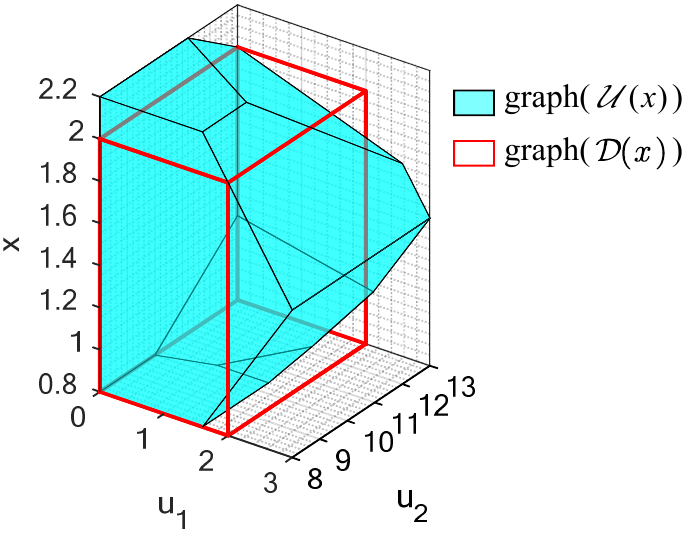}
	\caption{The graph of the DDUS $\mathcal{U}(x)$ and the dispatchable region $\mathcal{D}(x)$.}
	\label{fig:example:11}
\end{figure}

\begin{figure}[!htb]
	\centering
	\includegraphics[width=0.9\linewidth]{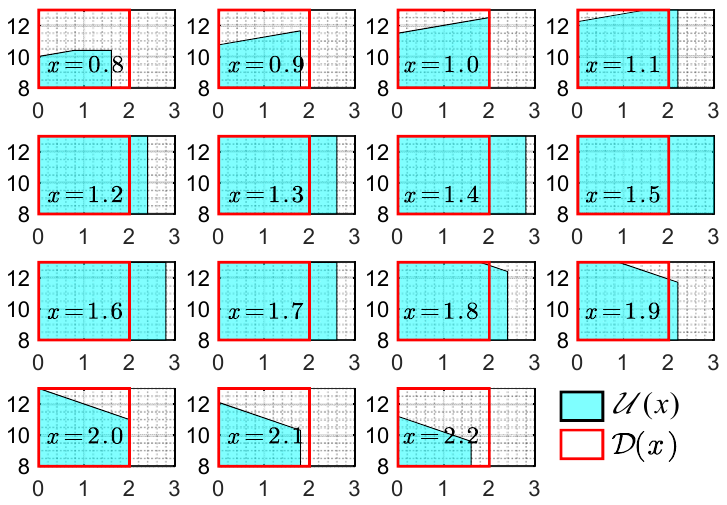}
	\caption{The DDUS $\mathcal{U}(x)$ and the dispatchable region $\mathcal{D}(x)$ under different values of $x$.}
	\label{fig:example:11-2}
\end{figure}

It is observed that when $1<x<2$, the dispatchable region $\mathcal{D}(x)$ is not able to cover the uncertainty set $\mathcal{U}(x)$. By recalling that the RFR $X_R$ contains all the $x$ that makes $\mathcal{U}(x)\subseteq \mathcal{D}(x)$ hold, the $X_R$ is obtained as
\begin{gather}
X_R=[0.8,1]\cup [2,2.2]
\end{gather}
and is obviously nonconvex.
\end{example}

The nonconvexity induced by polyhedral DDUS can be explained from two perspectives:

\textit{i)} Consider the TSRO problem \eqref{tsro} with polyhedral DDUS as follows
\begin{gather}
	\label{poly-DDUS}
	\mathcal{U}(x)=\left\{
	u| G(x)u\le g(x)
	\right\}
\end{gather}
where $G(x),g(x)$ are functions with respect to $x$. According to subsection \ref{subsec:partial_separa}, the $\mathcal{U}(x)$ is proved to be partially separable by considering $\mathcal{U}_{\rm fea}^{\rm sep}(x)$ in \eqref{sep:fea} and $\mathcal{U}_{\rm opt}^{\rm sep}(x)$ in \eqref{sep:opt}. Moreover, according to Theorem \ref{thm:eq}, the DDUS $\mathcal{U}(x)$ in TSRO problem \eqref{tsro} can be equivalently substituted by  $\mathcal{U}_{\rm fea}^{\rm sep}(x)$ and $\mathcal{U}_{\rm opt}^{\rm sep}(x)$ that are both completely separable. However, the coupling function of $\mathcal{U}_{\rm fea}^{\rm sep}(x)$, which is $\mathcal{C}_{\rm fea}(\cdot)$
in \eqref{sep:fea:3}, and the coupling function of $\mathcal{U}_{\rm opt}^{\rm sep}(x)$, which is $\mathcal{C}_{\rm opt}(\cdot)$ in \eqref{sep:opt:3}, are both nonconvex functions. Therefore, the convexity of the TSRO problem can not be ensured by applying Theorem \ref{thm:convex}.

\textit{ii)} Consider the polytope in the space of $(x,\alpha,u,y)$ as follows:
\begin{gather}
\label{ex-Y}
\left\{
(x,\alpha,u,y)|Y(x,u)\cap \left\{
y|c^{\mathsf{T}}y\le \alpha
\right\}
\right\}
\end{gather}
Denote the projection of \eqref{ex-Y} onto the space of $(x,\alpha,u)$ by
\begin{gather}
\label{psp2:eq1}
\left\{
(x,\alpha,u)|Hx+J\alpha+Ku\le M
\right\}
\end{gather}
where $H,J,K,M$ are constant matrices. According to Definition \ref{def:disp-reg}, \eqref{psp2:eq1} is the graph of $\mathcal{D}^{\rm ext}(x,\alpha)$. Then, constraint \eqref{eq-tsro:2} has equivalent formulations as in \eqref{psp2:eq2}.

\begin{subequations}
\label{psp2:eq2}
\begin{align}
\label{psp2:eq2:1}
\eqref{eq-tsro:2}
\overset{(i)}{=}&\left\{
(x,\alpha)| Hx+J\alpha+Ku\le M,
\forall u\in \mathcal{U}(x)
\right\}\\
\label{psp2:eq2:2}
\overset{(ii)}{=}&\left\{
(x,\alpha)|
\begin{array}{c}
h_j^{\mathsf{T}}x+J\alpha+\max_{u\in\mathcal{U}(x)}k_j^{\mathsf{T}}u\le m_j,\\
\forall j=1,\ldots,{\rm len}(M)
\end{array}
\right\}\\
\label{psp2:eq2:3}
\overset{(iii)}{=}&\left\{
(x,\alpha)|
\begin{array}{c}
h_j^{\mathsf{T}}x+J\alpha+\left\{
\begin{array}{c}
	\min_{\lambda_j}\ \lambda_j^{\mathsf{T}}g(x)\\
	\text{s.t.}\ G(x)^{\mathsf{T}}\lambda_j=k_j\\
	\lambda_j\ge \mathbf{0}
\end{array}
\right\}\\
\le m_j,\forall j=1,\ldots,{\rm len}(M)
\end{array}
\right\}\\
\label{psp2:eq2:4}
\overset{(iv)}{=}&\left\{
(x,\alpha)|
\begin{array}{c}
\left\{
\begin{array}{c}
h_j^{\mathsf{T}}x+J\alpha+\lambda_j^{\mathsf{T}}g(x)\le m_j\\
G(x)^{\mathsf{T}}\lambda_j=k_j\\
\lambda_j\ge \mathbf{0}
\end{array}
\right\}\neq \emptyset,\\
\forall j=1,\ldots,{\rm len}(M)
\end{array}
\right\}\\
\label{psp2:eq2:5}
\overset{(v)}{=}&\left\{
(x,\alpha)|
\underbrace{
\left\{
\begin{array}{c}
	Hx+J\alpha+\Lambda g(x)\le M\\
	\Lambda G(x)=K\\
	\Lambda\ge \mathbf{0}
\end{array}
\right\}}_{(*)}\neq \emptyset
\right\}
\end{align}
\end{subequations}

For equations \eqref{psp2:eq2}: in \textit{(ii)}, $h_j,k_j,m_j$ are the $j$th row of $H,K,M$, respectively, and ${\rm len}(M)$ denotes the number of rows in $M$; \textit{(iii)} is derived by recalling \eqref{poly-DDUS} and applying dual transformation to the inner maximization problem in \eqref{psp2:eq2:2} where $\lambda_j$ is the dual multiplier; \textit{(iv)} is obtained by equivalently removing the minimization operator in \eqref{psp2:eq2:3}; In \textit{(v)}, $\Lambda:=(\lambda_1,\ldots,\lambda_{\text{len}(M)})^{\mathsf{T}}$.

According to the transformation in \eqref{psp2:eq2}, constraint \eqref{eq-tsro:2} is equivalent to the projection of $(*)$ onto the space of $(x,\alpha)$ where $(*)$ is defined in \eqref{psp2:eq2:5}. By noting the bilinear term $\Lambda g(x)$ and $\Lambda G(x)$ in $(*)$, region $(*)$ may be nonconvex in the space of $(x,\alpha,\Lambda)$. Therefore, constraint \eqref{eq-tsro:2}, which is the projection of $(*)$, can also be nonconvex. When $G(x)$ and $g(x)$ are decision-independent constant matrices, region $(*)$ and its projection are guaranteed to be polyhedrons, justifying that constraint \eqref{eq-tsro:2} is convex in the case of DIUs.

\section{Algorithm for Robust Dispatch with Decision-Dependent Uncertainties} \label{sec_algo}

Existing literature have proposed some algorithms to solve specific robust dispatch problems with DDUs. However, the mechanism of DDU's influence on algorithms is still unclear. Compared to these works, this paper unfolds the separable property of DDU and then designs a generic solution algorithm for DDUs based on the coupling function. The proposed algorithm is not only practical but also inspiring by revealing the essential change when DDUs are considered. It is potential to improve existing algorithms by applying the theoretical results in this section.

Specifically, this section begins by reviewing two mainstream cutting-plane algorithms developed for TSRO problems and revealing their limitations in dealing with DDUs. Then, an improved solution strategy is proposed to solve the DDU-integrated TSRO problem.

\subsection{The Cutting-Plane Algorithms for TSRO Problem}
\label{subsec:rr}

\begin{figure}[!htb]
	\centering
	\includegraphics[width=0.9\linewidth]{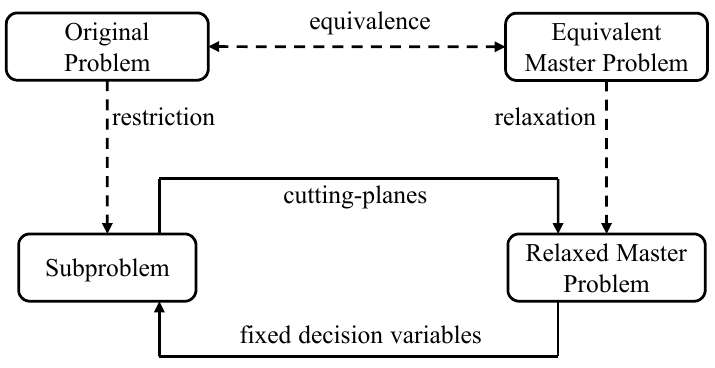}
	\caption{The principle of \textit{Restriction and Relaxation}.}
	\label{fig:r&r}
\end{figure}

Existing solution algorithms for TSRO problems, including the renowned Benders decomposition and C\&CG algorithm, follow the principle of \textit{Restriction and Relaxation (R\&R)}\cite{geoffrion1970elements1,geoffrion1970elements2}, see Fig. \ref{fig:r&r}. \textit{Restriction} is referred to fixing the values of some decision variables in an optimization problem and \textit{Relaxation} means dropping some of the constraints.

The basic idea of \textit{R\&R} is explained as follows. Denote by the TSRO problem to be solved as the original problem (OP). Without loss of generation, the OP is assumed to be a minimization problem. By applying an equivalent transformation to the OP, the so-called equivalent master problem (EMP) is derived. A subproblem (SP) is got by applying \textit{Restriction} to the OP and a relaxed master problem (RMP) is derived by applying \textit{Relaxation} to the EMP. The RMP and EMP share the same optimization objective.

The solution algorithms based on the principle of \textit{R\&R} are designed to iteratively solve the SP and the RMP. The solution to SP provides an upper bound of the optimum of OP and that of RMP provides a lower bound. Moreover, by solving the RMP, values of the fixed decision variables in the RMP are updated. By solving the SP, additional cutting-plane constraints are appended to RMP. The iteration terminates till the optimum of SP and RMP are close enough, implying the optimum to OP is obtained. Under the same solution framework of \textit{R\&R}, the Benders decomposition and the C\&CG algorithm mainly differ themselves in the formulations of cutting-planes, EMP, and RMP.

\subsubsection{The Benders Decomposition}
Recalling the sets $\Xi$ in \eqref{sep:fea:2} and $\Pi$ in \eqref{sep:opt:2}, the cutting-planes in Benders decomposition are designed as:
\begin{gather}
	\text{CP}^{\text{bd}}_{\rm fea}(\xi,u)=\left\{
	x|\xi^{\mathsf{T}}(b-Ax-Cu)\le 0
	\right\}\\
	\text{CP}^{\text{bd}}_{\rm opt}(\pi,u)=\left\{
	(x,\alpha)|\pi^{\mathsf{T}}(b-Ax-Cu)\le \alpha
	\right\}
\end{gather}
The $\text{CP}^{\text{bd}}_{\rm fea}(\xi,u)$ parameterized in $\xi\in \Xi$ and $u\in U$ is called the feasibility cut and is a half-plane in $\mathbb{R}^{n_x}$. The $\text{CP}^{\text{bd}}_{\rm opt}(\pi,u)$ parameterized in $\pi\in \Pi$ and $u\in U$ is the optimality cut which is a half-plane in $\mathbb{R}^{n_x+1}$. 

According to Theorem \ref{thm:eq}, the EMP in Benders decomposition which is surrogate to the original TSRO problem \eqref{tsro} can be formulated as follows:
\begin{subequations}
\label{bd:emp}
\begin{gather}
\label{bd:emp:1}
\min_{x\in X,\alpha\in \mathbb{R}^1}\ f(x)+\alpha\\
\label{bd:emp:2}
x\in \text{CP}^{\text{bd}}_{\rm fea}(\xi,u),\forall \xi \in \Xi,\forall u\in \mathcal{U}(x)\\
\label{bd:emp:3}
(x,\alpha)\in \text{CP}^{\text{bd}}_{\rm opt}(\pi,u),\forall \pi \in \Pi,\forall u\in \mathcal{U}(x)
\end{gather}
\end{subequations}

By applying partial enumeration to $\xi$ in $\Xi$, $\pi$ in $\Pi$, and $u$ in $\mathcal{U}(x)$, the Benders decomposition is intended to relax constraints \eqref{bd:emp:2}-\eqref{bd:emp:3} into linear cutting-planes. To be specific, if a tentative solution $x^j$ does not satisfy \eqref{bd:emp:2}, there exists $\xi^j\in\Xi$ and $u^j\in\mathcal{U}(x^j)$ such that $x^j \notin \text{CP}^{\text{bd}}_{\rm fea}(\xi^j,u^j)$. Then, in the RMP, a cutting-plane $x\in \text{CP}^{\text{bd}}_{\rm fea}(\xi^j,u^j)$ is formulated to approximate constraint \eqref{bd:emp:2}. Similarly, if a tentative solution $(x^j,\alpha^j)$ does not satisfy \eqref{bd:emp:3}, there exists $\pi^j\in \Pi$ and $u^j\in\mathcal{U}(x^j)$ such that $x^j\notin \text{CP}^{\text{bd}}_{\rm opt}(\pi^j,u^j)$. Then, in the RMP, a cutting-plane $(x,\alpha)\in \text{CP}^{\text{bd}}_{\rm opt}(\pi^j,u^j)$ is formulated to approximate constraint \eqref{bd:emp:3}. For details of Benders decomposition for TSRO, please refer to \cite{Benders,Jiang2012,bertsimas2013adaptive}.

\subsubsection{The C\&CG Algorithm}
Denote by $\text{CP}^{\text{ccg}}_{}(u)$ the cutting-plane in the C\&CG algorithm:
\begin{gather}
	\text{CP}^{\text{ccg}}_{}(u)=\left\{
	(x,\alpha)|Y(x,u)\cap \left\{y|c^{\mathsf{T}}y\le\alpha\right\}\neq\emptyset
	\right\}
\end{gather}
It is a half-plane in $\mathbb{R}^{n_x+1}$ parameterized in $u\in U$. When the C\&CG algorithm is applied to the TSRO problem \eqref{tsro}, the EMP is designed as:
\begin{subequations}
	\label{ccg:emp}
	\begin{gather}
		\label{ccg:emp:1}
		\min_{x\in X,\alpha\in \mathbb{R}^1}\ f(x)+\alpha\\
		\label{ccg:emp:2}
	(x,\alpha)\in \text{CP}^{\text{ccg}}_{}(u),\forall u\in\mathcal{U}(x)
	\end{gather}
\end{subequations}
The equivalence between the original TSRO problem \eqref{tsro} and the EMP \eqref{ccg:emp} is justified by recalling problem \eqref{eq-tsro}. By partially enumerating the $u\in \mathcal{U}(x)$ in constraint \eqref{ccg:emp:2}, the C\&CG algorithm is intended to relax constraint \eqref{ccg:emp:2} into linear cutting-planes. To be specific, if a tentative solution $(x^j,\alpha^j)$ does not satisfy constraint \eqref{ccg:emp:2}, there exists $u^j\in \mathcal{U}(x^j)$ such that $(x^j,\alpha^j)\notin \text{CP}^{\text{ccg}}(u^j)$. Then, in the RMP, a cutting-plane $(x,\alpha)\in \text{CP}^{\text{ccg}}_{}(u^j)$ is formulated to approximate constraint \eqref{ccg:emp:2}. For details of the C\&CG algorithm, please refer to \cite{zeng2013solving}.

\subsection{Limitations of Cutting-Plane Algorithms}
The inapplicability of the aforementioned cutting-plane algorithms to DDU-integrated TSRO problems is discussed from the following two perspectives.

\textit{i) Theoretical Perspective:} As has been discussed in Section \ref{sec:convexity}, DDUs may introduce nonconvexity to the OP \eqref{tsro}. Since the OP \eqref{tsro}, the EMP \eqref{bd:emp} in Benders decomposition, and the EMP \eqref{ccg:emp} in C\&CG algorithm are equivalent, the EMPs could also be nonconvex optimization problems. It is straightforward that the nonconvexity of EMPs, if any, must reside in the constraints of EMP. As has been stated in subsection \ref{subsec:rr}, the two cutting-plane algorithms are intended to relax and approximate the feasible region of EMP by a set of cutting-planes. When the feasible region of EMP becomes nonconvex due to DDUs, the RMPs formulated by cutting-plane constraints are not guaranteed to be a relaxation to the EMP, which may ruin the quality of the algorithm output.

\textit{ii) Physical Perspective:} Recalling the physical meaning of robust optimization, the solution to the TSRO problem hedges against the worst-case scenario within the uncertainty set. As has been discussed in subsection \ref{subsec:rr}, when a tentative solution $x^j$ is proved to be not content with the constraints in EMPs under scenario $u^j\in\mathcal{U}(x^j)$, additional cutting-planes parameterized in $u^j$ are generated to cope with the identified worst-case scenario. However, as the uncertainty set varies with decision $x$, $u^j$ may reside outside the uncertainty set when $x$ takes other values. Therefore, the cutting-planes parameterized in $u^j$ may bring over-conservatism to the RMP, which prevents the RMP from approximating the EMP.

Next, the potential consequences of applying conventional cutting-plane algorithms to solving DDUs-integrated TSRO problems are revealed and justified by illustrative examples.

\subsubsection{Failure in Robust Feasibility}
The RMP consists of cutting-planes is not guaranteed to be a relaxation to the EMP. The over-conservative cutting-plane maybe an empty set, making the RMP infeasible and interrupting the iterative algorithm. Once the decision maker observes an infeasible RMP, the OP would be misdiagnosed as infeasible. An example of the failure in robust feasibility is provided as follows. 
\begin{example}
\label{example12}
On the basis of Example \ref{example11}, consider the TSRO problem as follows:
\begin{subequations}
	\label{chp2:example:ro}
	\begin{gather}
		\min_{x\in [0.8,2.2]}\quad \vert x-1.5\vert\\
		\text{s.t.}\quad x\in X_R
	\end{gather}
\end{subequations}
According to Example \ref{example11}, the explicit formulation of $X_R$ is $[0.8,1]\cup [2,2.2]$. Therefore, the optimal solution to \eqref{chp2:example:ro} is $x^*=1$. 

When applying Benders decomposition to solving \eqref{chp2:example:ro}, the tentative solution in the first iteration round is $x^1=1.5$, which corresponds to $u^1=(3,8)^{\mathsf{T}}\in \mathcal{U}(x^1)$ and $\xi^1=(0,-1,0,-1,-1,0)^{\mathsf{T}}\in \Xi$ such that $x^1\notin \text{CP}_{\rm fea}^{\rm bd}(\xi^1,u^1)$. Then, a cutting-plane constraint $x\in \text{CP}_{\rm fea}^{\rm bd}(\xi^1,u^1)$ with explicit formulation as follows
\begin{gather}
	\label{chp2:example:benderscut}
	x\in\mathbb{R}^1\ :\ 
	\xi_1^1+\xi_2^1+\xi_3^1+\xi_4^1-u_1^1\xi_5^1+u_2^1\xi_6^1\le 0
\end{gather}
is appended to the EMP. However, it can be justified that constraint \eqref{chp2:example:benderscut} is an empty set that makes the RMP infeasible. On observing an infeasible RMP, the decision maker would be under the delusion that the original problem \eqref{chp2:example:ro} is infeasible.

When applying the C\&CG algorithm to solving \eqref{chp2:example:ro}, in the first iteration round, there exists $u^1=(3,8)^{\mathsf{T}}\in \mathcal{U}(x^1)$ such that $x^1\notin \text{CP}^{\rm ccg}(u^1)$. Then, a cutting-plane constraint $x\in \text{CP}^{\rm ccg}(u^1)$ with explicit formulation as follows
\begin{gather}
	\label{chp2:example:ccgcut}
	x\in\mathbb{R}^1\ :\ 
	\left\{
	\begin{array}{c}
		-1\le y_1\le 1\\
		-1\le y_2\le 1\\
		3\le y_1+y_2\le 8
	\end{array}
	\right\}\neq \emptyset
\end{gather}
is appended to the RMP. However, constraint \eqref{chp2:example:ccgcut} is obviously an empty set that would make the RMP infeasible. Similarly, the feasibility of problem \eqref{chp2:example:ro} would be misconceived by the decision maker.
\end{example}

\subsubsection{Failure in Robust Optimality}
The over-conservative cutting-planes in RMPs may shrink the feasibility region of the EMP, leading to a suboptimal solution to the OP, see the example as follows.
\begin{example}
\label{example13}
On the basis of Example \ref{example12}, consider the constraints of the second-stage problem as follows:
\begin{gather}
	\label{chp2:example:II:Y2}
	Y(x,u)=\left\{
	y\in\mathbb{R}^2\ |\ \begin{array}{c}
		-1\le y_1\le 1\quad \left(\xi_1,\xi_2\right)\\
		-1\le y_2\le 1\quad \left(\xi_3,\xi_4\right)\\
		u_1-0.5x\le y_1+y_2\quad \left(\xi_5\right)\\
		y_1+y_2\le u_2+0.5x\quad (\xi_6)
	\end{array} 
	\right\}
\end{gather}
The dispatchable region corresponds to $Y(x,u)$ in \eqref{chp2:example:II:Y2} is
\begin{gather}
	\label{chp2:example:dispreg}
	\mathcal{D}(x)=\left\{
	u\in\mathbb{R}^2\ |\ 
	\begin{array}{c}
		u_1\le 2+0.5x\\
		-u_2\le 2+0.5x\\
		u_1-u_2\le x
	\end{array}
	\right\}
\end{gather}
To derive the RFR, the DDUS $\mathcal{U}(x)$ in \eqref{chp2:eg:dduset}, the dispatchable region $\mathcal{D}(x)$ in \eqref{chp2:example:dispreg}, and their graphs are plotted in Fig. \ref{fig:example13}.

\begin{figure*}[!htb]
	\centering
	\subfloat[The graph of $\mathcal{U}(x)$.]{\label{fig:example13:a}
		\includegraphics[width=0.25\linewidth]{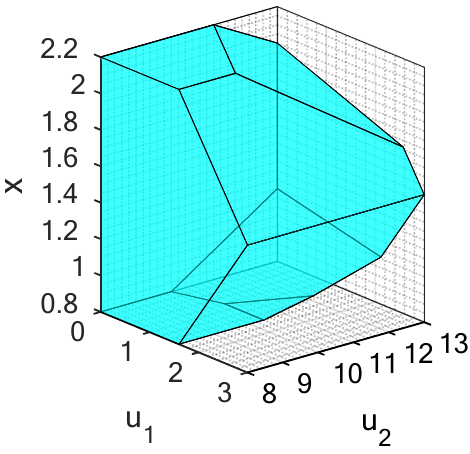}}
	\subfloat[$x=(0,1)^{\mathsf{T}}$]{\label{fig:example13:b}
		\includegraphics[width=0.25\linewidth]{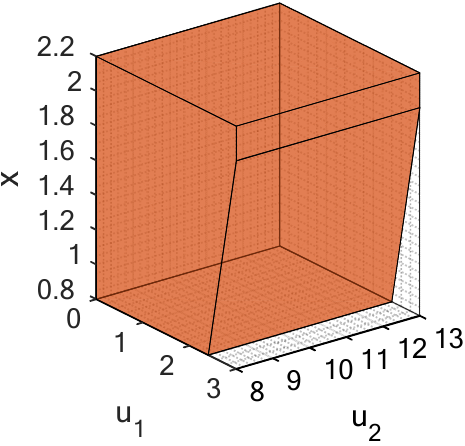}} 
	\subfloat[$x=(0.5,0.5)^{\mathsf{T}}$]{\label{fig:example13:c}
		\includegraphics[width=0.4\linewidth]{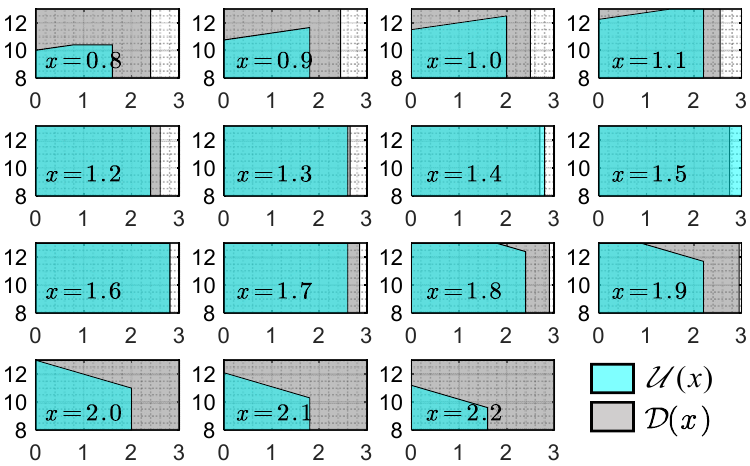}}
	\caption{Polyhedral decision-dependent uncertainty set and its partial separability.}
	\label{fig:example13} 
\end{figure*}

It can be inferred from Fig. \ref{fig:example13} that the explicit formulation of the RFR in this example is $X_R=[0.8,4/3]\cup [1.6,2.2]$. Therefore, the optimal solution in this example is $x^*=1.6$ and the optimum is 0.1.

When applying the Benders decomposition, the tentative solution in the first iteration round is $x^1=1.5$ and there exists $u^1=(3,8)^{\mathsf{T}}\in\mathcal{U}(x^1)$ and $\xi^1=(0,-1,0,-1,-1,0)^{\mathsf{T}}\in \Xi$ such that $x^1\notin \text{CP}_{\rm fea}^{\rm bd}(\xi^1,u^1)$. Then, a feasibility cutting-plane $x\in \text{CP}_{\rm fea}^{\rm bd}(\xi^1,u^1)$ with explicit formulation as follows
\begin{subequations}
\label{chp2:example:benderscut2}
\begin{align}
&	x\in\mathbb{R}^1\ :
\left\{
\begin{array}{c}
	\xi_1^1+\xi_2^1+\xi_3^1+\xi_4^1\\
	+(0.5x-u_1^1)\xi_5^1+(0.5x+u_2^1)\xi_6^1
\end{array}
\right\}\le 0\\
		\Leftrightarrow &x\in\mathbb{R}^1\ :\ 
		2\le x
\end{align}
\end{subequations}
is appended to the RMP. In the second iteration round, the RMP generates a tentative solution $x^2=2$ which is proved to be within the RFR $X_R$. Then, the iteration algorithm terminates and outturns a suboptimal solution $x^2=2$ with an objective value of 0.5.

When applying the C\&CG algorithm, the tentative solution in the first iteration round is also $x^1=1.5$ and there exists $u^1=(3,8)^{\mathsf{T}}\in\mathcal{U}(x^1)$ such that $x^1\notin \text{CP}^{\rm ccg}(u^1)$. Then, a cutting-plane $x\in \notin \text{CP}^{\rm ccg}(u^1)$ with explicit formulation as follows
\begin{subequations}
\label{example13:ccg}
\begin{align}
	&x\in\mathbb{R}^1:\ 
	\left\{
	\begin{array}{c}
		-1\le y_1\le 1\\
		-1\le y_2\le 1\\
		3-0.5x\le y_1+y_2\le 8+0.5 x
	\end{array}
	\right\}\neq \emptyset\\
	\Leftrightarrow\quad &
	x\in\mathbb{R}^1: x\ge 2
\end{align}
\end{subequations}
is appended to the RMP. Similarly, the RMP with constraint \eqref{example13:ccg} would outturn a suboptimal solution $x^2=2$ with an objective value of 0.5.
\end{example}

\subsection{Improved Solution Algorithm for DDUs}
In this subsection, an improved solution algorithm is proposed to solve the DDU-integrated TSRO problems based on the separability of DDUS.

\subsubsection{DDUS with Complete Separability} 

Without loss of generality, the DDUS $\mathcal{U}(x)$ is assumed to be separable as in \eqref{chp2:def:seperate}. Then, the enhanced C\&CG cut parameterized in $\xi\in \Xi$ is formulated as:
\begin{gather}
\text{EC}^{\rm ccg}(\xi):=\left\{
(x,\alpha,u)|\begin{array}{c}
(x,\alpha)\in \text{CP}^{\rm ccg}(u)\\
u^j=\mathcal{C}(\xi,x)
\end{array}
\right\}
\end{gather}

Then, the EMP \eqref{bd:emp} can be reformulated as
\begin{subequations}
\label{e-ccg-emp}
\begin{gather}
\label{e-ccg-emp:1}
\min_{x\in X,\alpha\in\mathbb{R}^1,u}\ f(x)+\alpha\\
\label{e-ccg-emp:2}
(x,\alpha,u)\in \text{EC}^{\rm ccg}(\xi),\forall \xi\in \Xi
\end{gather}
\end{subequations}
And by applying partial enumeration to $\xi\in \Xi$ in \eqref{e-ccg-emp:2}, the RMP is formulated as:
\begin{subequations}
	\label{e-ccg-rmp}
	\begin{gather}
		\label{e-ccg-rmp:1}
		\min_{}\ f(x)+\alpha\\
		\text{over:}\ x\in X,\alpha\in\mathbb{R}^1,u^j\in U\\
		\label{e-ccg-rmp:2}
		(x,\alpha,u^j)\in \text{EC}^{\rm ccg}(\xi^j),\forall j\
	\end{gather}
\end{subequations}
The RMP \eqref{e-ccg-rmp} provides a valid relaxation to the EMP \eqref{e-ccg-emp}. As the constraints in \eqref{e-ccg-rmp:2} accumulate, the RMP \eqref{e-ccg-rmp} is approaching the EMP \eqref{e-ccg-emp}.

\subsubsection{DDUS with Partial Separability}

The enhanced feasibility Benders cut parameterized in $\xi\in \Xi$ is formulated as:
\begin{gather}
\text{EC}_{\rm fea}^{\rm bd}(\xi):=
	\left\{
(x,u)|
\begin{array}{c}
x\in \textrm{CP}_{\rm fea}^{\rm bd}(\xi,u)\\
u=\mathcal{C}_{\rm fea}(\xi,x)
\end{array}
	\right\}
\end{gather}
The $\text{EC}_{\rm fea}^{\rm bd}(\xi)$ is a constraint in $\mathbb{R}^{n_x+n_u}$.
The enhanced optimality Benders cut is:
\begin{gather}
\text{EC}_{\rm opt}^{\rm bd}(\pi):=
\left\{
(x,\alpha,u)|
\begin{array}{c}
	(x,\alpha)\in \textrm{CP}_{\rm opt}^{\rm bd}(\pi,u)\\
	u=\mathcal{C}_{\rm opt}(\pi,x)
\end{array}
\right\}
\end{gather}
which is a constraint in $\mathbb{R}^{n_x+1+n_u}$.

Then the EMP can be formulated as
\begin{subequations}
\label{e-bd-emp}
\begin{gather}
\label{e-bd-emp:1}
\min\ f(x)+\alpha\\
\label{e-bd-emp:2}
\text{over:}\ x\in X,\alpha\in\mathbb{R}^1,u_{\rm fea},u_{\rm opt}\in U\\
\label{e-bd-emp:3}
(x,u_{\rm fea})\in \text{EC}_{\rm fea}^{\rm bd}(\xi),\forall \xi\in\Xi\\
\label{e-bd-emp:4}
(x,\alpha,u_{\rm opt})\in \text{EC}_{\rm opt}^{\rm bd}(\pi),\forall \pi\in\Pi
\end{gather}
\end{subequations}
By applying partial enumeration to $\xi\in \Xi$ in constraint \eqref{e-bd-emp:3} and $\pi\in \Pi$ in constraint \eqref{e-bd-emp:4}, the RMP is formulated into:
\begin{subequations}
\begin{gather}
\min\ f(x)+\alpha\\
\text{over:}\ x\in X,\alpha\in\mathbb{R}^1,u_{\rm fea}^j,u_{\rm opt}^k\in U\\
(x,u_{\rm fea}^j)\in \text{EC}_{\rm fea}^{\rm bd}(\xi^j),\forall j\\
(x,\alpha,u_{\rm opt}^k)\in \text{EC}_{\rm opt}^{\rm bd}(\pi^k),\forall k
\end{gather}
\end{subequations}

\section{Applications of Robust Dispatch with Decision-Dependent Uncertainties}
This section presents some applications of robust dispatch with DDUs including the frequency-constrained reserve allocation with DDUs of wind generators \cite{mypaperfreq}, robust dispatch considering the DDUs of demand response \cite{su2020robust}, and robust scheduling of virtual power plant (VPP) under DDU \cite{my2022robust}.

\subsection{Source Side: Frequency-Constrained Reserve Allocation of Wind Generators}
By synthetic inertia and droop control, RES units like wind generators can provide fast response for frequency regulation by reserve allocation. It means wind generators can be regarded as a kind of frequency regulation resource. Meanwhile, the reserve allocation problem of wind generators should consider the uncertainty of wind generation power. The decision of preserving power will influence the realization of the real available wind power, which result in the DDU.

The frequency-constrained reserve allocation problem of wind generators can be formulated by a TSRO problem. At stage one, the here-and-now decision $x$ is composed of the unit commitment decisions of synchronous generators and the energy dispatch and reserves allocation decisions of synchronous and wind generators. The reserve allocation affects the uncertainty of wind power. After the real wind power is observed, the stage-two decision $y$ including the redispatch of synchronous and wind generators is made for frequency regulation.

\textbf{Example \ref{example:2}} shows the DDUs of wind generators in different modes. To provide frequency reserve with a reliability guarantee, our work \cite{mypaperfreq} chooses the \textit{Delta} mode:
\begin{align} \label{Delta_Mode}
\text{\textit{Delta:}}\quad \left\{
\begin{array}{l}
R_{j,k} = \min\left\{P^{\rm mppt}_{j,k}, \lambda_{j,k} P^{\rm rate}_j \right\}\\
P^{\rm avail}_{j,k} = \max\left\{
P^{\rm mppt}_{j,k} - \lambda_{j,k} P^{\rm rate}_j, 0
\right\}
\end{array}
\right.
\end{align}
where $P^{\rm rate}_j$ is the rated power of wind generator $j$, $R_{j,k}, P^{\rm avail}_{j,k}, \lambda_{j,k}$, and $P^{\rm mppt}_{j,k}$ are the preserved power, the available power, the de-loading ratio, and the maximal available power of wind generator $j$ at time $k$, respectively. In \eqref{Delta_Mode}, $P^{\rm mppt}_{j,k}$ is a DIU variable solely determined by the uncertain wind speed. The preserved power $R_{j,k}$ and the available power $P^{\rm avail}_{j,k}$ are DDU variables dependent on the de-loading ratio decision $\lambda_{j,k}$.

As we impose robustness guarantee on $P^{\rm mppt}_{j,k} \ge \overline{R}_{j,k} := \lambda_{j,k} P^{\rm rate}_j$, the DDU model is further simplified by
\begin{subequations}
\begin{align}
\label{chp3:def_Xi0}
\overline{P}^{\rm avail}_{j,k}=P_{j,k}^{\rm mppt}-\overline{R}_{j,k}\ \forall j,k\\
\label{chp3:def_Xi1}
\underline{P}_{j,k}^{\rm mppt}\le P_{j,k}^{\rm mppt}\le \overline{P}_{j,k}^{\rm mppt}\ \forall j,k\\
\label{chp3:def_Xi2}
\sum\nolimits_{k}\vert P_{j,k}^{\rm mppt}- P_{j,k}^{\rm av} \vert/ P^{\rm h}_{j,k}\le \Gamma_j^{\rm T}\ \forall j\\
\label{chp3:def_Xi3}
\sum\nolimits_{j}\vert P_{j,k}^{\rm mppt}- P_{j,k}^{\rm av} \vert/ P^{\rm h}_{j,k}\le \Gamma_k^{\rm S}\ \forall k
\end{align}
\end{subequations}
where $P_{j,k}^{\rm av}$ and $P^{\rm h}_{j,k}$ are the expected value and fluctuation level of $P^{\rm mppt}_{j,k}$, respectively, $\Gamma_j^{\rm T}$ is the temporal robustness budget for wind generator $j$, and $\Gamma_k^{\rm S}$ is the spacial robustness budget for time $k$.

In the DDU model, the DDU variable is 
\begin{align}
u := \overline{P}^{\rm avail}_{j,k}\ \forall j,k
\end{align}
The variable affecting the DDU in the here-and-now decision $x$ is $\overline{R}_{j,k}\ \forall j,k$.

Based on the separability of DDUS, the auxiliary random variable is
\begin{align}
\xi := P_{j,k}^{\rm mppt}\ \forall j,k
\end{align}
the DIU support set $\Xi$ is defined by constraints \eqref{chp3:def_Xi1} - \eqref{chp3:def_Xi3}, and the coupling function $\mathcal{C}(\xi,x)$ is defined by equation \eqref{chp3:def_Xi0}. It is clear that the DDU set is \textbf{completely separable}. Therefore, the whole robust reserve allocation problem can be solved by the improved C\&CG algorithm proposed in Section \ref{sec_algo}.

\begin{figure}[!htb]
\centering
\includegraphics[width=0.8\linewidth]{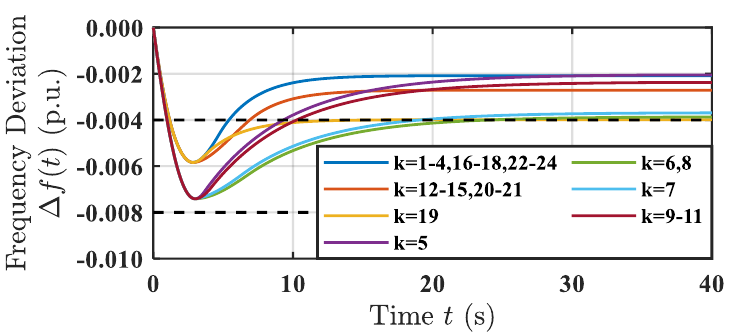}
\caption{Frequency performance of wind generator considering DDU \cite{mypaperfreq}.}
\label{fig:caseFrequencyDDU}
\end{figure}

\begin{figure}[!htb]
\centering
\includegraphics[width=0.8\linewidth]{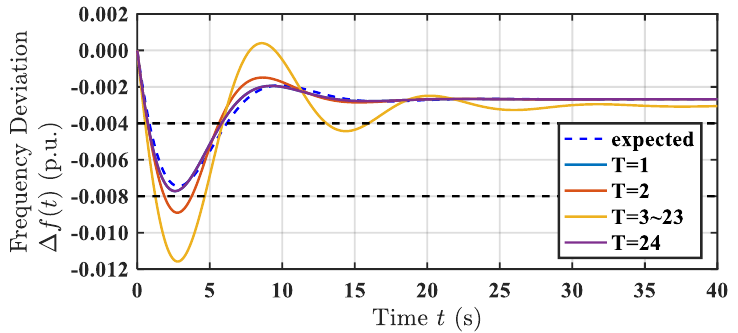}
\caption{Frequency performance of wind generator considering DIU \cite{mypaperfreq}.}
\label{fig:caseFrequencyDIU}
\end{figure}

Numerical results on a modified PJM 5-bus system are provided to verify the importance of DDU in robust dispatch. Figs. \ref{fig:caseFrequencyDDU} and \ref{fig:caseFrequencyDIU} show the frequency trajectories at each time considering DDU and DIU models, respectively. By considering the DDU model of wind power, the frequency deviation is still maintained in the expected region. On the contrary, if the wind power is modeled by the DIU set, it could not provide expected frequency support, leading to violations on the frequency nadir. The numerical comparison shows the superiority of modeling wind generation output as DDU rather than DIU.

Numerical tests on a real regional power grid of China with 57 conventional generators and 6 renewable energy power plants \cite{zhuo2019transmission} is carried out to verify the scalability of the proposed solution algorithm. To test the effect of RESs (including wind farms and PV stations) in the TSRO model with DDUs, two scenarios are considered as follows.

\begin{enumerate}
\item COMP. model: only synchronous generators provide frequency regulation service.

\item PPSD. model: synchronous generators and RESs all provide frequency regulation service.
\end{enumerate}

\begin{figure*}[t]
\centering
\includegraphics[width=0.95\linewidth]{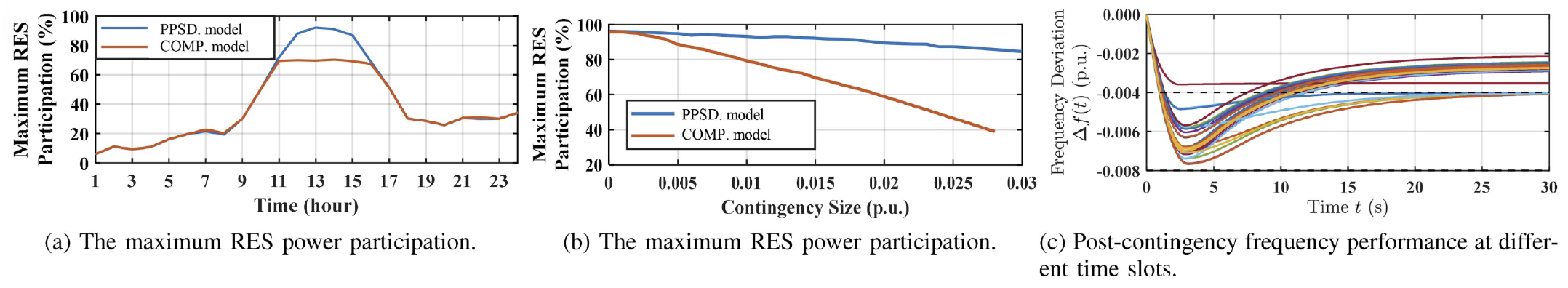}
\caption{Results of the regional power grid in China. Subfigures (a) and (b) refer to a deterministic model that maximizes RES power participation. Subfigure (c) refers to the robust model. \cite{mypaperfreq}}
\label{fig:caseFrequency_real_system}
\end{figure*}

The numerical results are shown in Fig. \ref{fig:caseFrequency_real_system}. In Fig. \ref{fig:caseFrequency_real_system}(a), during 11:00 - 16:00, compared to the PPSD. model, the proportion of RES generation power in the COMP. model is reduced to about 70\%. In Fig. \ref{fig:caseFrequency_real_system}(b), by taking the value of the contingency size from 0\% to 3\%, the penetration of RES decreases. It is obvious that the curve of PPSD. model drops much slower than that of COMP. model. Besides, the problem of the COMP. model becomes infeasible when the contingency size increases to 2.9\%. In Fig. \ref{fig:caseFrequency_real_system}(c), The post-contingency frequency dynamics at different time slots are plotted, verifying the robust feasibility. What's more, the computational time of the PPSD. model is 1914.22s, which is compatible with the day-ahead dispatch problem in real-world large power systems.

\subsection{Demand Side: Robust Dispatch with Demand Response}
Demand response is a promising solution to reducing operation costs and alleviating system risks by encouraging end-users to participate in system regulation. However, the demand itself is commonly stochastic. The decision of demand response may change probability distributions or bounds of demand. Therefore, the uncertainty of demand response belongs to DDU.

Our work \cite{su2020robust} studied the robust dispatch problem with demand response, which belongs to TSRO. At stage one, the here-and-now decision $x$ is composed of the nominal generations and reserve capacities of conventional units and the set-point of demand response $\bm{d}^0$. This problem concentrates on the DDU of demand response, which depends on the set-point $\bm{d}^0$. After the real loads are observed, the stage-two decision $y$, regulated generation outputs of conventional units, is made to maintain power balance and power flow constraints.

The loads themselves own the following uncertainty
\begin{align}
\mathcal{D}^e = \left\{ \bm{d} ~|~ -\delta_{d,j}^- \le d_j - d_j^e \le \delta_{d,j}^+,~ \forall j \right\}
\end{align}
where $\delta_{d,j}^-, \delta_{d,j}^+$ and $d_j^e$ are the upper and lower fluctuation, and the forecast value of load $j$.

By demand response, the forecast loads are changed to the set-point $\bm{d}^0$. Assume that the fluctuations of loads are proportionally amplified with the forecast load. Then the new uncertainty set is defined as
\begin{align} \label{DR_DDU_Set}
\mathcal{D}(\bm{d}^0) = \left\{ \bm{d} ~|~ -\frac{d_j^0}{d_j^e} \delta_{d,j}^- \le d_j - d_j^0 \le \frac{d_j^0}{d_j^e} \delta_{d,j}^+,~ \forall j \right\}
\end{align}
where $d_j^0 / d_j^e$ is the amplification ratio. Hence $\mathcal{D}(\bm{d}^0)$ is a DDUS depending on the here-and-now decision $\bm{d}^0$.

The DDUS of demand response belongs to Box DDUs in \textbf{Example \ref{example:1}}, which is \textbf{completely separable}. Hence the robust dispatch problem with demand response was solved by the improved C\&CG algorithm proposed in Section \ref{sec_algo}.

\begin{figure*}[t]
\centering
\includegraphics[width=0.95\linewidth]{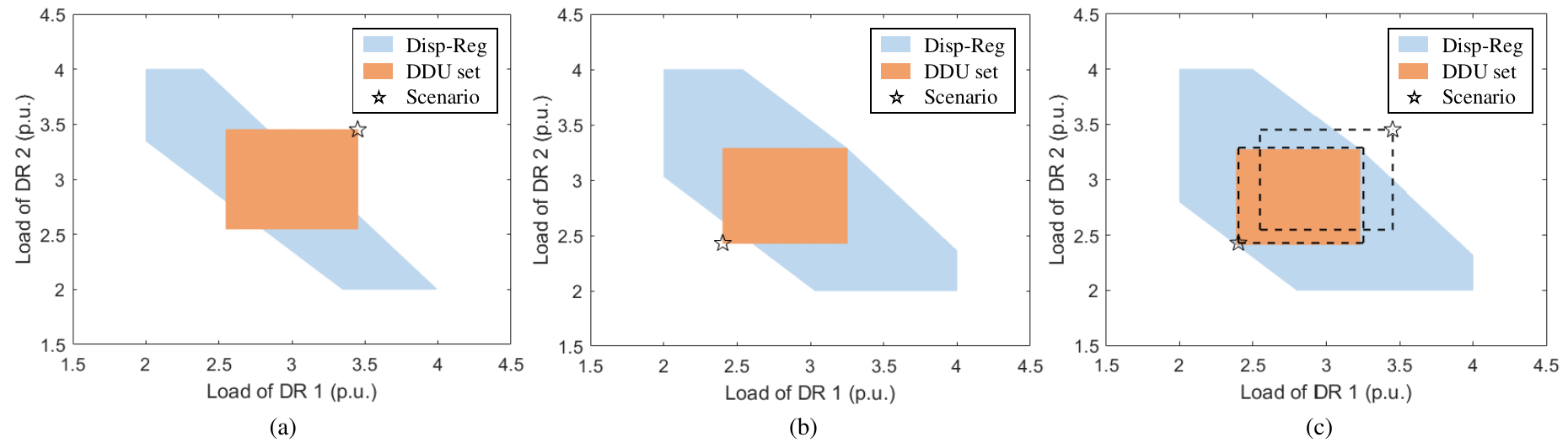}
\caption{Iteration procedure of improved C\&CG algorithm. (a) $k = 1$. (b) $k = 2$. (c) $k = 3$ \cite{su2020robust}.}
\label{fig:caseDemandDDU}
\end{figure*}

\begin{figure}[t]
\centering
\includegraphics[width=0.8\linewidth]{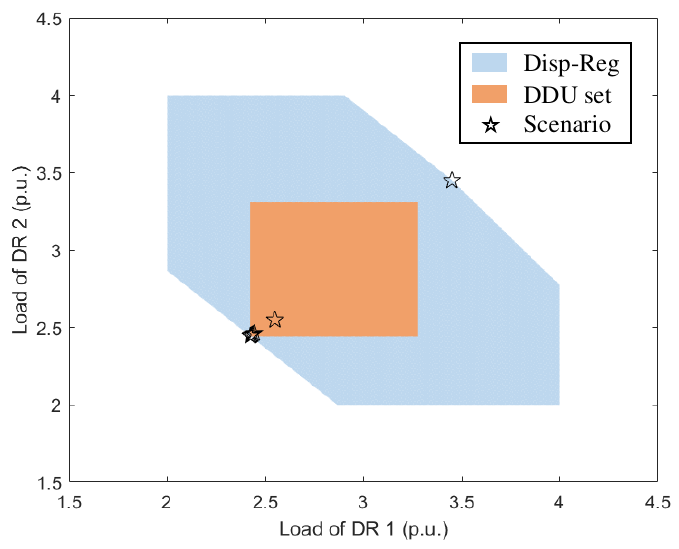}
\caption{Iteration result of standard C\&CG algorithm \cite{su2020robust}.}
\label{fig:caseDemandDIU}
\end{figure}

Our work \cite{su2020robust} illustrates the influence of the demand response DDU  and validates the effectiveness of the improved C\&CG algorithm in the IEEE 5-bus system with two loads participating in the demand response program. Fig. \ref{fig:caseDemandDDU} shows the iteration procedure of the improved C\&CG algorithm, where Disp-Reg means the dispatchable region in \textbf{Definition \ref{def:disp-reg}}. Fig. \ref{fig:caseDemandDDU} illustrates the bilateral matching between the DDUS and the dispatchable region. If the uncertainty is regarded to be decision-independent, the dispatch result may result in failures in robust feasibility or/and optimality. Fig. \ref{fig:caseDemandDIU} shows the iteration result of the standard C\&CG algorithm with the DIU model. The dispatchable region covers worst-case scenarios outside the uncertainty set, indicating the loss of robust optimality.

\subsection{Reserve Deployment: Robust Scheduling of Virtual Power Plant}
VPP can provide power systems with flexibility by aggregating renewable generation units, conventional power plants, energy storages, and flexible demands. Our work \cite{my2022robust} considers the scenario that a VPP participates in the day-ahead reserve market. From the perspective of the VPP, it does not know the real reserve deployment request in the day-ahead time scale. The VPP will provide a flexibility range and then the bulk grid will then transport a regulating signal varying within the flexibility range. For the VPP, the regulating signal is actually uncertain. Meanwhile, the uncertainty depends on the flexibility range it decides. Therefore, for the VPP, the regulating signal from the day-ahead reserve market belongs to DDU.

The robust scheduling problem of VPP can be formulated by a TSRO problem. At stage one, the VPP will make the here-and-now decision $x$ composed of the traded power, the upward/downward reserve capacity in the market, and the unit commitment of conventional power plants. The VPP should consider the DDU of regulating signal and the DIU of energy market price and wind generation power. At stage two, RESs including conventional power plants, energy storage units, and flexible demands are utilized to deal with the uncertainties.

The VPP focuses on the uncertain power exchange between it and the bulk grid $P^{\rm EXCH}_k$ defined as
\begin{gather}
\label{chp4:def:pexch}
P^{\rm EXCH}_k=P^{\rm E}_k+ P^{\rm R+}_k-P^{\rm R-}_k,\forall k\in \mathcal{T}
\end{gather}
where $P^{\rm E}_k$ is the decision of power traded in the energy market at time
period $k$, $P^{\rm R+}_k$ and $P^{\rm R-}_k$ are the uncertain upward and downward regulating signals to the VPP at time period $k$, respectively.

The uncertainties of upward and downward regulating signals are described by the following constraints
\begin{subequations}
\label{chp4:pR}
\begin{gather}
\label{chp4:pR:1}
0\le P^{\rm R+}_k\le P_k^{\rm RC+},\forall k\in \mathcal{T}\\
\label{chp4:pR:2}
0\le P^{\rm R-}_k\le P_k^{\rm RC-},\forall k\in \mathcal{T}\\
\label{chp4:pR:3}
\sum\nolimits_{k\in \mathcal{T}} P^{\rm R+}_k \le \sum\nolimits_{k\in \mathcal{T}}{sig_k^+}P^{\rm RC+}_k\\
\label{chp4:pR:4}
\sum\nolimits_{k\in \mathcal{T}} P^{\rm R-}_k \le \sum\nolimits_{k\in \mathcal{T}}{sig_k^{-}}P^{\rm RC-}_k
\end{gather}
\end{subequations}
where $sig_k^+$ and $sig_k^-$ are the average normalized upward/downward reserve energy called on to provide at time period $k$, $P_k^{\rm RC+}$ and $P_k^{\rm RC-}$ are the decisions of upward and downward reserve capacities, respectively.

The uncertainty set of $P^{\rm EXCH}_k$ is formulated by constraints \eqref{chp4:def:pexch} and \eqref{chp4:pR}. Since the set depends on the stage-one decisions $P^{\rm E}_k, P_k^{\rm RC+}$ and $P_k^{\rm RC-}$, it belong to DDUS. According to Section \ref{sec2}, this complex DDUS is \textbf{partially separable}. From Section \ref{sec:convexity}, TSRO problems with partially separable DDUS could be nonconvex. The proposed improved C\&CG algorithm in Section \ref{sec_algo} may face failures in robust feasibility or/and optimality. Section \ref{sec_algo} also proposes an improved Benders decomposition algorithm to deal with partially separable DDUS.

Our work \cite{my2022robust} verifies the effectiveness of the improved Benders decomposition algorithm in solving the robust VPP scheduling problem. The standard C\&CG algorithm is utilized to solve the same VPP scheduling problem. Table \ref{tab:chp4:standardCCG} shows the outcomes of both algorithms. The net profit by the standard C\&CG algorithm is much smaller than that by the improved Benders decomposition algorithm. This is because, in the CCG algorithm, the feasibility cut is directly generated by the worst-case uncertainty realization, ignoring the fact that the uncertainty set varies with decisions. What's more, the improved Benders decomposition algorithm requires fewer iterations and less solution time, displaying the efficiency of the Benders-dual cutting plane.

\begin{table}[!htb]
\caption{Comparison Between Standard C\&CG Algorithm and Improved Benders Decomposition \cite{my2022robust}}
\label{tab:chp4:standardCCG}
\centering
\begin{tabular}{cccc}
\toprule
Algorithm		& Net Profit [\$]  & Iteration & Solution Time [s] \\
\midrule
Standard C\&CG              & 7737.81    &25      &3106.18  \\
Improved Benders        & 7832.99    &11      &854.42   \\
\bottomrule           
\end{tabular}
\end{table}

In this section, we give some applications to show what is DDU in power system dispatch and how to deal with DDU. In fact, the proposed theorems and algorithms still work in other RESs, for instance, photovoltaics, energy storage units, and market participators.

\section{Conclusion}
This paper provides a systematic framework for two-stage robust dispatch with DDUs, shedding light on the fundamental correlations and distinctions between DDUs and DIUs. 
The structural properties of DDUs are explored first through the concept of separability, with rigorous definition, proved existence, and illustrative examples. It is revealed that a DDUS with an arbitrary set-valued mapping formulation can be decomposed into a coupling function parameterized in decisions and a decision-irrelevant uncertainty set. 
Subsequently, combined with the region-based flexibility characterization, robust dispatch with DDUs is equivalently formulated as the bilateral matching between the DDUS and the dispatchable region or its variant. Compared with the unilateral matching under DIUs, the bi-directional matching facilitated by DDUs endows more system flexibility, albeit potentially introducing non-convexity to the robust dispatch problem. It is proved that the non-convexity, if any, is attributed to the non-convex coupling function. 
Then an improved algorithm is proposed to precisely and efficiently solve the generic DDU-integrated TSRO dispatch, as a variant of the iterative algorithm within the framework of restriction and relaxation. Unlike traditional primal or dual cutting-planes, the enhanced cuts designed based on the separability of DDUS autonomously adapt to the convexity or non-convexity of the problem. 
Finally, applications to robust dispatch problems considering the DDUs of wind generators, demand response, and virtual power plants are introduced, which emphasize the importance of considering DDUs and verify the effectiveness of proposed algorithms to deal with DDUs.

It should be noted that though this paper focuses on the two-stage robust optimization, its key idea of DDU could still work in other optimization problems, e.g., multi-stage robust optimization and two/multi-stage distributionally robust optimization. It is expected that this study could provide a better understanding of the characterizations of DDUs and the algorithms for two-stage robust dispatch with DDU, and then inspire more theoretical and practical works concerning DDUs.

\appendix
\section*{Proof of Lemma \ref{lemma:1}}
\label{app-A}
Denote by $\mathcal{U}(x)={U}^{\rm sub}\cap \mathcal{U}^{\rm sup}(x),\forall x\in X$ and $$\mathcal{U}^{\dagger}(x)=\left\{
u=\mathcal{C}(u^{'},x)|u^{'}\in {U}^{\rm sub}\right\},\forall x\in X.$$ Next, we would like to justify the equivalence of $\mathcal{U}(x)$ and $\mathcal{U}^{\dagger}(x)$.

First, according to the property of $\mathcal{C}(\cdot)$,  we have $\mathcal{U}^{\dagger}(x)\subseteq \mathcal{U}(x),\forall x\in X$. Moreover, for any 
$u\in\mathcal{U}(x)$, there is $u=\mathcal{C}(u,x)$. Set $u^{'}=u$, then there exists $u^{'}\in {U}^{\rm sub}$ such that $u=\mathcal{C}(u^{'},x)$, implying that $u\in\mathcal{U}^{\dagger}(x)$. Therefore, $\mathcal{U}(x)\subseteq \mathcal{U}^{\dagger}(x),\forall x\in X$. This completes the proof that $\mathcal{U}(x)=\mathcal{U}^{\dagger}(x),\forall x\in X$.

\section*{Proof of Theorem \ref{thm:eq}}
\label{app-B}
\subsubsection*{Assertion a)}
First, denote by $R(x,u)$ the optimal value of the optimization problem and its dual formulation as follows:
\begin{subequations}
	\label{chp2:def:R:fea}
	\begin{align}
		\label{chp2:def:R:fea:1}
		R(x,u)&\overset{(i)}{:=}
		\left\{
		\begin{array}{ll}
			\min_{y,s}\quad &\mathbf{1}^{\mathsf{T}}s\\
			\text{s.t.}\quad &Ax+By+Cu\le b+s\\
			&s\ge \textbf{0},y\ge \textbf{0}
		\end{array}
		\right\}
		\\
		\label{chp2:def:R:fea:2}
		&\overset{(ii)}{=}\left\{
		\begin{array}{ll}
			\max_{\xi}\quad & \xi^{\mathsf{T}}(b-Ax-Cu)\\
			\text{s.t.}\quad & \xi\in\Xi
		\end{array}
		\right\}
	\end{align}
\end{subequations}
where $(i)$ is to define value function $R(x,u)$ and $(ii)$ is from the dual transformation of the minimization problem in \eqref{chp2:def:R:fea:1}. It is easy to verify that for any $x\in X$ and $u\in U$, $Y(x,u)\neq\emptyset \Leftrightarrow R(x,u)\le 0$. Denote by $\mathcal{U}_{\rm fea}^{*}(x)$ the set-valued map as follows
\begin{align}
\label{def:ufea*}
\mathcal{U}_{\rm fea}^{*}(x)=\arg_{u}
\left\{
\begin{array}{ll}
	\max\nolimits_{u,\xi}&\xi^{\mathsf{T}}(b-Ax-Cu)\\
	\text{s.t.}&u\in\mathcal{U}(x),\xi\in \Xi
\end{array}
\right\},\forall x\in X
\end{align}
For any $x\in X$:
\begin{subequations}
\label{xr:eq:v2}
\begin{align}
&Y(x,u)\neq \emptyset,\forall u\in\mathcal{U}(x)\\
\overset{(i)}{\Leftrightarrow}\quad & R(x,u)\le 0,\forall u\in\mathcal{U}(x)\\
\overset{(ii)}{\Leftrightarrow}\quad &\left\{
\begin{array}{ll}
	\max\nolimits_{u,\xi}\ &\xi^{\mathsf{T}}(b-Ax-Cu)\\
	\text{s.t.}\ &u\in\mathcal{U}(x),\xi\in \Xi
\end{array}
\right\}\le 0\\
\overset{(iii)}{\Leftrightarrow}\quad & \left\{
\begin{array}{ll}
	\max\nolimits_{u,\xi}\ &\xi^{\mathsf{T}}(b-Ax-Cu)\\
	\text{s.t.}\ &u\in\mathcal{U}_{\rm fea}^{*}(x),\xi\in \Xi
\end{array}
\right\}\le 0\\
\overset{(iv)}{\Leftrightarrow}\quad & R(x,u)\le 0,\forall u\in \mathcal{U}_{\rm fea}^*(x)\\
\overset{(v)}{\Leftrightarrow}\quad & Y(x,u)\neq \emptyset,\forall u\in \mathcal{U}_{\rm fea}^*(x)
\end{align}
\end{subequations}
where $(i)$ and $(v)$ are according to the equivalent formulation of $Y(x,u)\neq \emptyset$, $(ii)$ and $(iv)$ are according to the definition of $R(x,u)$, $(iii)$ is according to the definition of $\mathcal{U}_{\rm fea}^*(x)$.

According to the definition of $\mathcal{U}_{\rm fea}^*(x)$ in \eqref{def:ufea*} and that of $\mathcal{U}_{\rm fea}^{\rm sep}$ in \eqref{sep:fea}, the following relationship holds:
\begin{gather}
	\label{relation}
	\mathcal{U}_{\rm fea}^*(x)\subseteq \mathcal{U}_{\rm fea}^{\rm sep}(x)\subseteq \mathcal{U}(x),\forall x\in X
\end{gather}
Then, robust feasibility set $X_R$ is with equivalent formulations as follows:
\begin{subequations}
	\label{chp2:XR:eq}
	\begin{align}
		X_R&=\left\{x\in X\ |\ Y(x,u)\neq \emptyset,\forall u\in \mathcal{U}(x)\right\}\\
		&\overset{(i)}{=}\left\{x\in X\ |\ Y(x,u)\neq \emptyset,\forall u\in \mathcal{U}_{\rm fea}^*(x)\right\}\\
		&\overset{(ii)}{=}\left\{x\in X\ |\ Y(x,u)\neq \emptyset,\forall u\in \mathcal{U}_{\rm fea}^{\rm sep}(x)\right\}
	\end{align}
\end{subequations}
where $(i)$ is from the relationships in \eqref{xr:eq:v2} and $(ii)$ is from \eqref{relation}.

\subsubsection*{Assertion b)}
First, denote by $\mathcal{U}_{\rm opt}^{*}(x)$ the optimal solution set of $u$ for optimization problem $S(x)$:
\begin{subequations}
	\label{def:uopt*}
	\begin{align}
		\label{def:uopt*:1}
		\mathcal{U}_{\rm opt}^{*}(x)\overset{(i)}{:=}&\arg_u\ S(x),\forall x\in X\\
		\label{def:uopt*:2}
			\overset{(ii)}{=}&\arg_u\left\{
		\begin{array}{ll}
			\max_{u,\pi}&\pi^{\mathsf{T}}(b-Ax-Cu)\\
			\text{s.t.}&u\in\mathcal{U}(x),\pi\in \Pi\\
		\end{array}
		\right\},\forall x\in X
	\end{align}
\end{subequations}
where $(i)$ is the definition of $\mathcal{U}_{\rm opt}^{*}(x)$ and $(ii)$ is according to the dual transformation on the inner minimization problem in $S(x)$. According to the definition of $\mathcal{U}_{\rm opt}^{\rm sep}(x)$ in \eqref{sep:opt} and the $\mathcal{U}_{\rm opt}^{*}(x)$ in \eqref{def:uopt*}, the following relationships are derived:
\begin{gather}
	\label{chp2:nested2}
	\mathcal{U}_{\rm opt}^*(x)\subseteq \mathcal{U}_{\rm opt}^{\rm sep}(x)\subseteq \mathcal{U}(x),\forall x\in X
\end{gather}
Then, assertion b) is justified as follows
\begin{subequations}
\begin{align}
S(x)\overset{(i)}{:=}&\max_{u\in\mathcal{U}(x)}\min_{y\in Y(x,u)}\ c^{\mathsf{T}}y,\forall x\in X\\
\overset{(ii)}{=}&\max_{u\in\mathcal{U}_{\rm opt}^*(x)}\min_{y\in Y(x,u)}\ c^{\mathsf{T}}y,\forall x\in X\\
\overset{(iii)}{=}&\max_{u\in\mathcal{U}_{\rm opt}^{\rm sep}(x)}\min_{y\in Y(x,u)}\ c^{\mathsf{T}}y,\forall x\in X
\end{align}
\end{subequations}
where $(i)$ is from the definition of $S(x)$ in \eqref{tsro:3}, $(ii)$ is from the definition of $\mathcal{U}_{\rm opt}^*(x)$ in \eqref{def:uopt*}, and $(iii)$ is according to the relationships in \eqref{chp2:nested2}.

\section*{Proof of Lemma \ref{lemma:diu:convex}}
\label{app-C}
Consider set $\Gamma$ as follows:
\begin{gather}
	\Gamma:=
\cap_{u\in\mathcal{U}^0}\left\{
(x,\alpha,y)|Y(x,u) \cap \{y|c^{\mathsf{T}}y\le \alpha\}
\neq \emptyset
\right\}
\end{gather}
Since $Y(\cdot)$ and $\mathcal{U}^0$ are both polyhedrons, $\Gamma$ is equivalent to:
\begin{gather}
	\label{app-C:eq:1}
	\Gamma=
	\cap_{u\in\text{vert}(\mathcal{U}^0)}\left\{
	(x,\alpha,y)|Y(x,u) \cap \{y|c^{\mathsf{T}}y\le \alpha\}
	\neq \emptyset
	\right\}
\end{gather}
According to \eqref{app-C:eq:1}, $\Gamma$ is the intersection of a finite number of polyhedrons. Therefore, $\Gamma$ is a polytope in the space of $(x,\alpha,y)$. 

By projecting $\Gamma$ onto the subspace of $(x,\alpha)$, constraint \eqref{eq-tsro:2} is derived. Since the projection of a polyhedron onto its subspace is still a polyhedron, \eqref{eq-tsro:2} is proved to be a polyhedron. This completes the proof that \eqref{eq-tsro:2} is convex.

\section*{Proof of Theorem \ref{thm:convex}}
\label{app-D}
Let Assumption \ref{assp-tsro} hold. Assume that $\mathcal{T}$ is convex.

First, define optimization problem $R(x,\alpha,u^{'})$ as follows:
\begin{subequations}
	\label{chp2:def:R}
	\begin{align}
		\label{chp2:def:R:1}
		R(x,\alpha,\xi)\triangleq  \min\nolimits_{y,s}\quad &g(y,s)\\
		\label{chp2:def:R:2}
		\text{s.t.}\quad & Ax+By+\mathcal{T}(\xi,x)\le b+s_1\\
		\label{chp2:def:R:3}
		&y+ s_2\ge \mathbf{0}\\
		\label{chp2:def:R:4}
		&c^{\mathsf{T}}y\le \alpha+s_3\\
		\label{chp2:def:R:5}
		&s_1,s_2,s_3\ge \mathbf{0}
	\end{align}
\end{subequations}
where the objective function $g(y,s)$ is:
\begin{gather}
	g(y,s)=\mathbf{1}^{\mathsf{T}}s_1+\mathbf{1}^{\mathsf{T}}s_2+\mathbf{1}^{\mathsf{T}}s_3
\end{gather}
The objective function $g(\cdot)$ is linear and the feasible region \eqref{chp2:def:R:2}-\eqref{chp2:def:R:5} consist of a polyhedron.

Next, we would like to justify that $R(x,\alpha,\xi)$ is convex with respect to $x$, $\alpha$, and $\xi$. For any $\lambda\in [0,1]$, $(x^1,\alpha^1,\xi^{1})$, and $(x^2,\alpha^2,\xi^{2})$, denote by $(y^1,s^1)$ the optimal solution of $R(x^1,\alpha^1,\xi^1)$ and $(y^2,s^2)$ the optimal solution of $R(x^2,\alpha^2,\xi^2)$. Therefore, we have
\begin{subequations}
	\begin{gather}
		\left(x^1,\alpha^1,\xi^{1},y^1,s^1
		\right)\in \eqref{chp2:def:R:2}\text{-}\eqref{chp2:def:R:5}\\
		\left(x^2,\alpha^2,\xi^{2},y^2,s^2
		\right)\in \eqref{chp2:def:R:2}\text{-}\eqref{chp2:def:R:5}
	\end{gather}
\end{subequations}
Since $\mathcal{T}(\cdot)$ is a convex function, constraints \eqref{chp2:def:R:2}-\eqref{chp2:def:R:5} is convex with respect to $(x,\alpha,\xi,y,s)$. Therefore, we have
\begin{gather}
	\left(
	\begin{array}{c}
		\lambda x^1+(1-\lambda)x^2\\
		\lambda\alpha^1+(1-\lambda)\alpha^2\\
		\lambda \xi^{1}+(1-\lambda)\xi^{2}\\
		\lambda y^1+(1-\lambda)y^2\\
		\lambda s^1+(1-\lambda)s^2
	\end{array}
	\right)\in \eqref{chp2:def:R:2}\text{-}\eqref{chp2:def:R:5}
\end{gather}
which indicates that $\lambda y^1+(1-\lambda)y^2,\lambda s^1+(1-\lambda)s^2$ is the feasible solution to problem
$$
R\left(\lambda x^1+(1-\lambda)x^2,\lambda\alpha^1+(1-\lambda)\alpha^2,\lambda \xi^{1}+(1-\lambda)\xi^{2}\right)
$$
Therefore, the following relationships hold.
\begin{subequations}
	\label{chp2:R:relation}
	\begin{align}
		&\lambda R(x^1,\alpha^1,\xi^{1})+(1-\lambda)R(x^2,\alpha^2,\xi^{2})\\
		\overset{(a)}{=}\quad &\lambda g(y^1,s^1)+(1-\lambda)g(y^2,s^2)\\
		\overset{(b)}{=}\quad&g\left(
		\lambda y^1+(1-\lambda)y^2,\lambda s^1+(1-\lambda)s^2
		\right)\\
		\overset{(c)}{\ge}\quad&
		R\left(\lambda x^1+(1-\lambda)x^2,\lambda\alpha^1+(1-\lambda)\alpha^2,\lambda \xi^{1}+(1-\lambda)\xi^{2}\right)
	\end{align}
\end{subequations}

In \eqref{chp2:R:relation}, equation (a) is because $(y^1,s^1)$ is optimal to $R(x^1,\alpha^1,\xi^1)$ and $(y^2,s^2)$ is optimal to $R(x^2,\alpha^2,\xi^2)$; Equation (b) is because function $g(\cdot)$ is linear; Inequality (c) is because $\lambda y^1+(1-\lambda)y^2,\lambda s^1+(1-\lambda)s^2$ is the feasible solution to the minimization problem $R\left(\lambda x^1+(1-\lambda)x^2,\lambda\alpha^1+(1-\lambda)\alpha^2,\lambda \xi^{1}+(1-\lambda)\xi^{2}\right)$.

Equations \eqref{chp2:R:relation} justify that $R(x,\alpha,\xi)$ is a convex function with respect to $x$, $\alpha$, and $\xi$.

According to the definition of $\mathcal{D}^{\rm ext}(x,\alpha)$ in \eqref{def-ex-dis-reg} and the definition of $R(\cdot)$ in \eqref{chp2:def:R}, we have:
\begin{subequations}
	\begin{align}
		&(x,\alpha)\in \eqref{eq-tsro:2}\\
		\Leftrightarrow\quad &
		Y(x,\mathcal{C}(\xi,x))
		\cap \{
		y|c^{\mathsf{T}}y\le \alpha
		\}
		\neq \emptyset,\forall \xi\in \Xi\\
		\Leftrightarrow\quad &
		\left\{
		\begin{array}{c}
			Ax+By+\mathcal{T}(\xi,x)\le b\\
			y\ge \mathbf{0}\\
			c^{\mathsf{T}}y\le \alpha
		\end{array}
		\right\}\neq \emptyset,\forall \xi\in \Xi\\
		\Leftrightarrow\quad & R(x,\alpha,\xi)\le 0,\forall \xi\in \Xi
	\end{align}
\end{subequations}

Therefore, constraint \eqref{eq-tsro:2} equals to
\begin{gather}
	\eqref{eq-tsro:2}=\left\{
	x\in \mathbb{R}^{n_x},\alpha\in \mathbb{R}^1\ |\ 
	R(x,\alpha,\xi)\le 0,\forall \xi\in \Xi
	\right\}
\end{gather}

Next, we prove \eqref{eq-tsro:2} is convex by contradiction. Suppose there exist $(x^1,\alpha^1),(x^2,\alpha^2)\in \eqref{eq-tsro:2}$ and $\lambda\in [0,1]$ such that $(\lambda x^1+(1-\lambda)x^2,\lambda \alpha^1+(1-\lambda)\alpha^2)\notin \eqref{eq-tsro:2}$. This implies there exists $\xi^{*}\in \Xi$ such that
\begin{gather}
	\label{chp2:proof:ineq}
	R\left(\lambda x^1+(1-\lambda)x^2,\lambda \alpha^1+(1-\lambda)\alpha^2,{\xi}^{*}\right)>0
\end{gather}

Since $(x^1,\alpha^1),(x^2,\alpha^2)\in \eqref{eq-tsro:2}$, we have
$R(x^1,\alpha^1,\xi^{*})\le 0$ and $R(x^2,\alpha^2,\xi^{*})\le 0$. Since $R(\cdot)$ is convex, we have
\begin{subequations}
\begin{align}
	&R\left(
	\lambda x^1+(1-\lambda)x^2,\lambda \alpha^1+(1-\lambda)\alpha^2,\xi^{*}
	\right)\\
	\le\quad &\lambda R(x^1,\alpha^1,\xi^{*})+(1-\lambda)R(x^2,\alpha^2,\xi^{*})\\
	\le\quad & 0
\end{align}
\end{subequations}
which contradicts with \eqref{chp2:proof:ineq}. This completes the proof that \eqref{eq-tsro:2} is convex.

\ifCLASSOPTIONcaptionsoff
  \newpage
\fi
\bibliographystyle{IEEEtran}
\bibliography{IEEEabrv,refs}

\end{document}